\documentclass[a4paper,reqno]{amsart}

\pdfoutput=1

\usepackage[utf8]{inputenc}
\usepackage[T1]{fontenc}
\usepackage{lmodern}
\usepackage{amsthm, amssymb, amsmath, amsfonts, mathrsfs}

\usepackage{xcolor}

\usepackage{cancel}

\usepackage{microtype}

\usepackage[pagebackref,colorlinks=true,pdfpagemode=none,urlcolor=blue,
linkcolor=blue,citecolor=blue]{hyperref}

 \usepackage{graphics,tikz}

\usepackage{relsize}

\usepackage{amsmath,amsfonts,amssymb,amsthm}
\usepackage{amsthm, amssymb, amsmath, amsfonts, mathrsfs}

\usepackage{mathrsfs,fge}
\usepackage{MnSymbol}
\usepackage{scalerel} 

\usepackage{color}
\usepackage{accents}

\usepackage[normalem]{ulem}
\usepackage{bbm}

\usepackage{cleveref}

\definecolor{labelkey}{gray}{.8}
\definecolor{refkey}{gray}{.8}

\definecolor{darkred}{rgb}{0.9,0.1,0.1}
\definecolor{darkgreen}{rgb}{0,0.5,0}

\numberwithin{equation}{section}

\newcommand{\bbZ}{{\mathbb Z}}
\newcommand{\bbP}{{\mathbb P}}
\newcommand{\bbR}{{\mathbb R}}

\newcommand{\bbT}{{\mathbb T}}
\newcommand{\eps}{\varepsilon}

\newcommand{\si}{\sigma}
\newcommand{\Om}{\Omega}
\newcommand{\om}{\omega}
\newcommand{\omm}{\omega_{\rm min}}
\newcommand{\omM}{\omega_{\rm max}}

\newcommand{\ga}{\gamma}
\newcommand{\la}{\lambda}
\newcommand{\al}{\alpha}
\newcommand{\bbE}{\mathbb E}

\newcommand{\cal}{\mathcal}

\newtheorem{df}{Definition}[section]
\newtheorem{remark}[df]{Remark}
\newtheorem{lm}[df]{Lemma}
\newtheorem{lemma}[df]{Lemma}
\newtheorem{prop}[df]{Proposition}
\newtheorem{thm}[df]{Theorem}

\newcommand{\blue}{\textcolor{blue}}

\newcommand{\commentout}[1]{{}}

\makeatother

\begin{document}


\title[Asymptotic Scattering]{Asymptotic Scattering by Poissonian Thermostats}

\author{Tomasz Komorowski  }
 \thanks{The corresponding author: Tomasz Komorowski, e-mail:{ \tt tkomorowski@impan.pl}} 

\author{Stefano Olla}

\address[Tomasz Komorowski -  the corresponding author]{Institute of Mathematics,
  Polish Academy Of Sciences, ul. \'Sniadeckich 8, 00-636 Warsaw, Poland,
e-mail:{ \tt tkomorowski@impan.pl}}

\address[Stefano Olla]{CEREMADE, UMR-CNRS,
  Universit\'e de Paris Dauphine, PSL Research University and  Institute
  Universitaire de France and Gran Sasso Science Institute (GSSI),
e-mail:{ \tt olla@ceremade.dauphine.fr}}

\maketitle

\begin{abstract}
 In the present paper we consider an infinite  chain of harmonic
  oscillators coupled with a
Poisson thermostat attached at a point.
The kinetic limit for the energy density of the chain, given by the Wigner
distribution, satisfies a transport equation outside the
thermostat location. A boundary condition
 emerges at this site, which describes the
reflection-transmission-scattering of the wave energy 
{ scattered} {off} by the thermostat. Formulas for the respective
coefficients are obtained. Unlike the case of the Langevin
thermostat studied in \cite{kors}, 
the Poissonian thermostat scattering {generates} in {the    limit } a continuous cloud
of waves of frequencies different from {that} of the incident wave.  
 \end{abstract}

\section{Introduction}
\label{intro}

{ {In the present paper} we consider a one-dimensional  infinite
  chain of harmonic oscillators, with a thermostat attached a point.
  The thermostat, maintained {at} a fixed   temperature $T$, is   usually modelled, at the microscopic level, by 
some stochastic process: e.g. by the Langevin stochastic dynamics, or
by  the renewal of velocities  at random times 
with Gaussian distributed velocities of variance $T$. The latter   represents the interaction with an infinitely extended reservoir 
of independent particles in equilibrium at temperature $T$ and uniform density. 
A natural question arises to describe the effect of a
thermostat on the wave energy density propagation in  the system in a
large space-time scale  limit. In the paper we 
  investigate this issue in the case of  the kinetic (hyperbolic) space-time 
scaling. This question has been studied for   a Langevin
thermostat in the recent article \cite{kors}. The goal of this paper is to find out how other classes of
thermostats, in particular of the Poisson type,  influence the energy
transport in the chain in the kinetic limit.}

 {More specifically,  consider an infinite one-dimensional chain of harmonic oscillators,
where particles are labelled by {the elements of the integer
  lattice}  $\mathbb Z$.} {The chain is
 coupled}  with a   thermostat acting on the particle labelled $0$.
The thermostat is modelled by a random mechanism depending on two parameters:
$\gamma> 0$, { describing its strength}, and  $ \mu\ge 1/2$, {whose
role is more technical as it decribes an interpolation between Poisson
and Gaussian mechanisms.} At random times determined by a Poisson process of intensity
$\gamma\mu$, the velocity $p_0$ of the particle $0$ is changed to
$$
p_0' =\left(1-\frac 1\mu\right) p_0 + \frac{\sqrt{2\mu -1}}{\mu} \tilde p,
$$
where $\tilde p$ is a centered Gaussian random variable with variance $T$
(the temperature of the thermostat).
{The case $\mu = 1/2$ corresponds to a  velocity flip  from
  $p_0\mapsto -p_0$ at Poisson random
times, $\mu=1$  ensures
complete renewal of $p_0$,  replacing it at those times by a $\mathcal N(0,T)$ random
variable $\tilde p$.} {Letting $\mu\to\infty$ {the process described in
the foregoing} converges to the Langevin thermostat considered in
\cite{kors}(cf. \eqref{therm-sde1}).
In this sense the parameter $\mu$ allows to {interpolate between
various models of  thermostats: starting from
the random flip process ($\mu=1/2$), through the
simple complete Poisson renewal  ($\mu=1$) and ending up at the
Langevin thermostat  ($\mu=+\infty$).}

{In the case $\mu = 1/2$ (the random velocity flip) the
energy of the chain is conserved and there is no thermalization. 
  On the other hand, when $\mu> 1/2$ the Gaussian distribution $\mathcal N(0,T)$
is the only stationary measure that is asymptotically
  stable for the process associated with the thermostat and the
  thermalization of the chain at
  temperature $T$ occurs.}

{To describe the energy density  distribution  in the space and
  frequency domain   we use the Wigner distribution. 
When there is no thermostat  present, the limit  of the Wigner distribution,
under the hyperbolic scaling,
is the solution of a simple transport equation. It describes the
evolution of the density of 
phonons, 
travelling independently of each other, with the group velocity
$\omega'(k)/2\pi$ corresponding to the phonon of wavenumber $k$. Here
$\om(k)$ is the dispersion relation of the harmonic chain and
a wavenumber $k$ belongs to $\bbT$ - the unit torus.}
{Taking into account the presence of the thermostat the respective
  limit, see   \eqref{010304} below, can be decomposed into the parts
that, besides  the aforementioned free energy transport, correspond to the production, absorption,  scattering,
transmission and reflection of a phonon. }
More precisely, we show that {when the dispersion relation is unimodal,
see Section \ref{sec:dynamics-the-main} for a precise definition}, 
in the scaling limit, the thermostat at temperature $T>0$
{ and corresponding to $\mu\ge 1/2$}
enforces the following reflection-transmission (and production) conditions at $x=0$:
phonons of wavenumber $\ell$  are generated {  at the rate $p_{\rm abs} \fgeeszett(\ell)T$
and an incoming $\ell$-phonon, arriving with velocity $\bar\om'(\ell)$,  is transmitted with probability $p_+(\ell)$,
reflected with probability $p_-(\ell)$, scattered, as a $k$-phonon, with the outgoing
velocity $\bar\om'(k)$, according to the scattering kernel $ \fgeeszett(\ell)p_{\rm sc}(k)$, and absorbed with probability $p_{\rm abs}\fgeeszett(\ell)$, see formulas  \eqref{033110} below. }
These coefficients are {non-negative}, depend on $\omega(\cdot)$,
the parameters $\ga>0$ and  $\mu\ge 1/2$, and satisfy 
$$
 p_+(\ell) + p_-(\ell) +p_{\rm abs}\fgeeszett(\ell) +
 \fgeeszett(\ell)\int_{\bbT}p_{\rm sc}(k)d k= 1,\quad {\ell\in
\bbT}.
$$
  Coefficients $p_\pm(\ell), \fgeeszett(\ell)$ do not depend on $\mu$. The coefficient
  $p_{\rm abs}$ is independent of $\ell$ and
  for  $\mu\to+\infty$,  $p_{\rm abs}\to 1$  and $p_{\rm sc}(k)\to 0$.
With such boundary conditions
the thermal equilibrium Wigner function $W(t,x,k) = T$ is a {stationary}
solution of the transport equation {for any $\mu>1/2$}. }

{  Our result covers also the random flip of sign of $p_0$, i.e. $\mu = 1/2$.
  In this case there is no absorbtion of phonons: $p_{\rm abs} = 0$, and
  $\int_{\bbT}p_{\rm sc}(k)dk= 1$, i.e. all the energy that is not
  transmitted or reflected at the same frequency is {scattered at various} frequencies. 
}

{The thermostat {corresponding to a finite value of $\mu$} plays a role
of  a  ``scatterer" of 
time-varying strength.} At the {macroscopic scale}  a 
wave incident on the thermostat   produces
reflected and transmitted waves at all frequencies. 
{  This is in stark constrast with the case of the Langevin thermostat ($\mu=+\infty$)
considered in \cite{kors}, where, after the scaling limit, the reflected and transmitted waves 
are of the same frequency as the incident wave ($p_{\rm sc}(k)=0$). }

{  Similarly to  \cite{kors} the presence of oscillatory integrals, responsible for the damping mechanism, presents the  difficulty
of the model and is dealt with using  the Laplace transform of the Wigner distribution. An additional difficulty lies in the fact that, contrary to  \cite{kors}, the noise  appearing in the dynamics
\eqref{eq:bas2} is multiplicative (rather than additive as in ibid.), which makes the computations much less explicit.}



{Introducing a rarefied random scattering in the bulk,
  in the same fashion as in \cite{BOS}, should
lead to a similar transport equation with a linear scattering term, 
without {modifying} the conditions at the interface with the
thermostat. Analogous case for the Langevin thermostat has been
considered in \cite{KORS2}.
}


 {
{\bf Acknowledgement}. TK   acknowledges the
support of the  Polish  National Science Centre:
Grant No.  2020/37/B/ST1/00426, 
SO by the French Agence Nationale Recherche grant LSD ANR-15-CE40-0020-01.
}

\section{Preliminaries  and formulation of the main result}
\label{sec:dynamics-the-main}

\subsection{Notation}

We use the notation $\bbT_a=[-a/2,a/2]$ for the torus of size $a>0$,
with identified endpoints. In particular for $a=1$ we write $\bbT$
instead of $\bbT_1$. We shall also write
$\bbT_+:=[k\in\bbT:\,0<k<1/2]$ and $\bbT_-:=[k\in\bbT:\,-1/2<k<0]$.

The Fourier transform of  a square integrable sequence $(\alpha_x) $ and the inverse Fourier
transform of $\hat \alpha\in L^2(\bbT)$ are defined as
  \begin{equation}
  \label{fourier}
  \hat \alpha (k)=\sum_{x\in\bbZ} \alpha_x \exp\{-2\pi ixk\}, ~~
\alpha_x=\int_{\bbT} \hat \alpha (k) \exp\{2\pi ixk\} dk, \quad x\in \bbZ,~~k\in\bbT.
\end{equation} 
Suppose that $f,g\in L^1[0,+\infty)$. 
Their convolution, also belonging to $L^1[0,+\infty)$,   is given by
$$
f\star g(t):=\int_0^tf(t-s)g(s)ds,\quad t\in[0,+\infty)
$$ 
By $f^{\star,k}$ we denote the $k$-times convolution of $f$ with
itself, i.e. $f^{\star,1}:=f$, $f^{\star,k+1}:=f\star f^{\star,k}$, $k\ge1$.
We let $f^{\star,0}\star g:=g$. We denote by
$$
\tilde f(\la)=\int_0^{+\infty}e^{-\la t}f(t)dt,\quad {\rm Re}\,\la>0,
$$
the Laplace transform of $f$. 
We also use the notation 
\begin{equation}
\label{asb}
(a\star b)_y=\sum_{y'\in\bbZ}a_{y-y'}b_{y'},\quad y\in\bbZ
\end{equation}
for the convolution of two absolutely summable sequences $(a_y)_{y\in\bbZ}$, $(b_y)_{y\in\bbZ}$.

Given a function $G(x,k)$, we denote by $\tilde G:\bbR\times\bbZ\to\mathbb C$,  $\hat
G:\bbR\times\bbT\to\mathbb C$ the Fourier transforms of $G$
in the $k$ and $x$ variables, respectively,
\begin{equation}
\label{022201-21}
\begin{split}
&\tilde G(x,y):=\int_{\bbT}e^{-2\pi i ky}G(x,k)dk,\quad
(x,y)\in\bbR\times\bbZ,\\
&
\hat G(\eta,k):=\int_{\bbR}e^{-2\pi i \eta x}G(x,k)dx,\quad (\eta,k)\in\bbR\times \bbT.
\end{split}
\end{equation}

Let us denote by
${\cal A}$ the Banach space obtained as the completion of ${\cal S}(\bbR\times\bbT)$ in the norm
\begin{equation}
\label{norm-ta01}
\| G\|_{{\cal A}}:=\int_{\bbR}\sup_{k\in\bbT} |\hat G(\eta,k)|d\eta
\end{equation}
and by ${\cal A}'$ its dual.

\subsection{Poisson type thermostat} 
\label{sec2.3}
The stochastic process
describing a thermostat is a jump process, whose  generator 
is given by
\begin{equation}
\label{L-gen}
L_{\mu,\ga}f ({\frak p}):=\frac{\ga \mu}{\sqrt{2\pi
    T}}\int_{\bbR}\left[f\left(\left(1-\frac{1}{\mu}\right){\frak p}+\rho(\mu)
    \tilde{\frak p}\right)-f ({\frak
    p})\right]\exp\left\{-\frac{\tilde{\frak
      p}^2}{2T}\right\}d\tilde{\frak p},\quad f\in B_b(\bbR).
\end{equation}
Here $B_b(\bbR)$ denotes the space of all bounded and Borel
measurable functions, $T,\ga>0$, $\mu\ge 1/2$ and 
\begin{equation}
\label{rho-mu}
\rho(\mu):=\frac{\sqrt{2\mu-1}}{\mu} .
\end{equation}
It is easy to verify that the Gaussian measure ${\cal N}(0,T)$ is invariant
under the dynamics of the process.
{  In the case $\mu = 1/2$ 
Gaussian measure ${\cal N}(0,T')$ is invariant for each $T'\ge0$.}

The process $({\frak p}_t)_{t\ge0}$   can be also described using the It\^o stochastic differential equation,
with a noise corresponding to a Poisson jump process, see e.g. \cite[Chapter V]{protter},
\begin{equation}
\label{therm-sde}
\begin{split}
&d{\frak p}(t) =\left(\tilde p(t-)-\frac{1}{\mu}{\frak
    p}(t-)\right)dN(\ga\mu t),\quad t\ge0,\\
&
{\frak p}(0)=\tilde p_0.
\end{split}
\end{equation}
Here  $\left(N(t)\right)_{t\ge0}$ is a Poisson process of
intensity $1$ defined over some probability space $(\Om,{\cal
   F},\bbP)$  and $\left(\tilde p(t)\right)_{t\ge0}$ is given by
\begin{equation}
\label{tp}
{\tilde
   p}(t):=\rho(\mu)\tilde p_{N'(\ga\mu t)},
\end{equation}
where $N'(t)=N(t)+1$.
We suppose that
  $(\tilde p_j)_{j\ge0}$ are i.i.d. ${\cal N}(0,T)$  random variables
  over $(\Om,{\cal  F},\bbP)$.

The process $\left({\tilde
   p}(t)\right)_{t\ge0}$ is Levy stationary  and 
\begin{equation}
\label{010906-20}
\begin{aligned}
&\bbE\, {\tilde
   p}(t)=0,\\
&\bbE\left[ {\tilde
   p}(t) {\tilde
   p}(t')\right]=\frac{2\mu-1}{\mu^2}e^{-\ga\mu|t-t'|}T,\quad t,t'\ge0.
\end{aligned}
\end{equation}  
From equation \eqref{therm-sde} we can see that in case $\mu=1$ we
have
$
{\tilde
   p}(t)=\tilde p_{N'(\ga t)}$, $t\ge0.$
 On the other hand, after a simple calculation,
 from \eqref{L-gen}, we conclude that for any $f\in
C^2(\bbR)$ 
\begin{equation}
\label{012511-20}
\lim_{\mu\to+\infty}L_{\mu,\ga}f ({\frak p})=L_{\infty,\ga} f ({\frak
  p}):=\ga T\exp\left\{\frac{{\frak
      p}^2}{2T}\right\}\frac{d}{d {\frak
      p}}\left(\exp\left\{-\frac{{\frak
      p}^2}{2T}\right\}\frac{df({\frak
      p})}{d {\frak
      p}}\right).
\end{equation}
The termostat correspong to $\mu=+\infty$ can be therefore identified
with the Langevin thermostat at temperature $T$, whose dynamics is described by the
It\^o stochastic differential equation, with an additive Gaussian white noise $dw(t)$:
\begin{equation}
\label{therm-sde1}
\begin{split}
&d{\frak p}(t) =-\ga {\frak
    p}(t)dt+\sqrt{2\ga T}dw(t),\quad t\ge0,\\
&
{\frak p}(0)=\tilde p_0.
\end{split}
\end{equation}
This case has been considered in \cite{kors}.

\subsection{Harmonic chain coupled with a point thermostat}

{We  couple the
particle with label $y=0$ with a thermostat described in Section \ref{sec2.3}.
Then the   
dynamics of the chain, with a stochastic source at $y=0$,  {is} governed by
\begin{eqnarray}
&&\dot{\frak q}_y(t)={\frak p}_y(t),
\label{eq:bas2}\\
&& d{\frak p}_y(t) = -(\alpha\star {\frak q}(t))_ydt+
\delta_{0,y}\left(\rho(\mu)\tilde p_{N(\ga\mu t)}-\frac{1}{\mu}{\frak p}_y(t-)\right)dN(\ga\mu t),\quad y\in\bbZ.\nonumber
\end{eqnarray}
The convolution operator $\star$ is defined in \eqref{asb}. The
coupling constants $(\al_y)_{y\in\bbZ}$ are even  $\al_{-y}=\al_y$ for all
$y\in\bbZ$ and real valued. In addition, we assume that they decay
exponentially, i.e. there exists $C>0$ so that 
\begin{equation}
\label{011601-21}
|\alpha_y|\le Ce^{-|y|/C}, \hbox{ for all $y\in \bbZ$,}
\end{equation}
and  
\begin{equation}
\label{alk}
\hat \al(k):=\sum_{y}\al_y\exp\left\{-2\pi i ky\right\}>0,\quad k\in\bbT_*:=\bbT\setminus\{0\}.
\end{equation}
Estimate \eqref{011601-21} in particular implies that
$\hat\alpha\in C^{\infty}(\bbT)$. By
 $({\frak q},{\frak p})=\big({\frak p}_y,{\frak q}_y\big)_{y\in\bbZ}$
 we denote the entire momentum-position configuration.  Equation \eqref{eq:bas2}
possesses a unique (mild) cadlag solution taking values in the space of square summable sequences
$({\frak q},{\frak p})$, see e.g. \cite[Section 9.4]{PZ}.}

\subsubsection {The dispersion relation and its basic properties} 

\label{sec2.4.1}

{Define the dispersion relation
\begin{equation}\label{mar2602}
 \om(k):=\sqrt{\hat \alpha (k)},\quad k\in\bbT.
\end{equation}
In light of \eqref{alk}, it is $C^\infty$ regular when
$\hat\al(0)>0$. If, on the other hand $\hat\al(0)=0$, the dispersion relation
 is a continuous function
on $\bbT$ belonging to 
$C^\infty(\bbT_*)$, with the derivatives possessing one sided
limits at $k=0$.
{ The typical examples are provided by the \emph{acoustic chains},
  where $\omega(k) \sim |k|$ for $k\sim 0$, and the \emph{optical chains} where
$\omega'(k) \sim k$ for $k\sim 0$.}
We assume also that $\om$ is {\em unimodal},
i.e. it is increasing on $[0,1/2]$. Denote its 
unique minimum, attained at $k=0$, by~$\omm\ge 0$ and its unique maximum, 
attained at~$k=1/2$, by $\omM$. The two branches of the inverse of
$\om(\cdot)$ are denoted by 
$\om_+:[\omm,\omM]\to[0,1/2]$ and $\om_-=-\om_+$.  }

\subsubsection{The wave-function}

{Define the complex valued wave function 
\begin{equation}
\label{011307}
\psi_y(t) := (\tilde{\om} \star{\frak q}(t))_ y + i{\frak p_y}(t).
\end{equation}
Here $\big(\tilde \om_y\big)_{y\in\bbZ}$ is the inverse
 Fourier transform of the  dispersion relation  $\om(k)$.}
The square of the wave function $|\psi_y(t)|^2$ describes the local
energy of the chain at time $t$. The Fourier transform of  $\big(\psi_y(t)\big)_{y\in\bbZ}$  is given by 
\begin{equation}
\label{011307a}
\hat\psi(t,k) := \om(k) \hat {\frak q}(t,k) + i\hat{\frak
  p}(t,k),\quad k\in\bbT.
\end{equation}
We have
$$
\hat{\frak
  p}\left(t,k\right)=\frac{1}{2i}[\hat\psi(t,k)-\hat\psi^*(t,-k)]\quad\mbox{and
}\qquad 
~~{\frak p}_0(t)=\int_{\bbT} {\rm Im}\,\hat\psi(t,k) dk.
$$
Using \eqref{eq:bas2}, it is easy to check that the wave function evolves according to  
\begin{equation}
\begin{split}
 \label{basic:sde:2aas}
 d\hat\psi(t,k) &= -i\om(k)\hat\psi(t,k)
dt
 +i\left(\tilde p(t-)-\frac{1}{\mu}{\frak p}_0(t-)\right)dN(\ga\mu t).
\end{split}
 \end{equation}

\subsubsection{The initial conditions}

\label{sec2.4.3}

{ Assume that for a given (small) value of the parameter $\eps>0$, the initial wave function is distributed randomly,
according to a Borel
probability measure~$\mu_\eps$ on the space of square summable
configurations. We suppose that
\begin{equation}
\label{0001}
\sup_{\eps\in(0,1)}\sum_{y\in\bbZ}\eps\langle|\psi_y|^2\rangle_{\mu_\eps}=\sup_{\eps\in(0,1)}\eps\langle \|\hat \psi\|^2_{L^2(\bbT)}\rangle_{\mu_\eps}
<\infty.
\end{equation}
Here $\langle\cdot\rangle_{\mu_\eps}$ denotes the expectation with
respect to $\mu_\eps$. Assumption \eqref{0001} guarantees that the
energy density  per unit length on the macroscopic scale $x\sim\eps y$
stays finite, as $\eps\to0+$.}

{In addition, to simplify somewhat our ensuing calculations, we will also assume that 
\begin{equation}
\label{null}
\langle\hat\psi(k)\hat\psi(\ell) \rangle_{\mu_\eps}=0,\quad k,\ell\in\bbT,
\end{equation}
The above hypothesis is of purely technical nature. It   can be
replaced by somewhat more general assumption that $
\langle\hat\psi(k)\hat\psi(\ell) \rangle_{\mu_\eps} \sim 0$, as $\eps
\to 0$, with no significant change in the main line our argument. However the
calculations would become more involved.
Later on   we shall also assume some  additional hypothesis,
see   \eqref{011812aa} below.}

\subsubsection{The Wigner distributions}

\label{sec2.4.4}

{Denote the
rescaled wave function $\psi^{(\eps)}_y(t)=\psi_y(t/\eps)$ and its
Fourier transform $\hat
  \psi^{(\eps)}(t,k)$.  The   (averaged)   Wigner
distributions  $ W^{(\eps)}_{\pm}(t)$ and $ Y^{(\eps)}_{\pm}(t)$ 
are defined by their action on a test function 
$G \in {\cal S}(\bbR\times\bbT)$:}
\begin{equation}
\label{wigner1}
\begin{split}
&\langle G,W^{(\eps)}_{\pm}(t)\rangle=\int_{\bbT\times\bbR} \widehat{
  W}_{\varepsilon,\pm}(t,\eta,k)\hat G^*(\eta,k)d\eta dk,\\
&\langle G,Y^{(\eps)}_{\pm}(t)\rangle=\int_{\bbT\times\bbR} \widehat{
  Y}_{\varepsilon,\pm}(t,\eta,k)\hat G^*(\eta,k)d\eta dk ,
\end{split}
\end{equation}
where
   \begin{align}
 \label{wigner1a}
&\widehat{ W}_{\eps,\pm}(t,\eta,k):=\frac{\eps}{2}
\bbE \left[\left(\hat \psi^{(\eps)}\right)^*\left(t,\pm k-\frac{\eps\eta}{2}\right)\hat
  \psi^{(\eps)}\left(t,\pm k+\frac{\eps \eta}{2}\right)\right], \nonumber
\\
&
\widehat{  Y}_{\eps,+}(t,\eta,k)
:=\frac{\eps}{2}  \bbE \left[\hat
  \psi^{(\eps)}\left(t,k+\frac{\eps \eta}{2}\right)\hat \psi^{(\eps)}\left(t,-k+\frac{\eps
      \eta}{2}\right)\right],\\
&
\widehat{  Y}_{\eps,-}(t,\eta,k)
:=\frac{\eps}{2}  \bbE \left[\left(\hat
  \psi^{(\eps)}\right)^\star\left(t,k-\frac{\eps \eta}{2}\right)\left(\hat \psi^{(\eps)}\right)^\star\left(t,-k-\frac{\eps
      \eta}{2}\right)\right],\quad (\eta,k)\in\bbT_{2/\eps}\times\bbT\nonumber
\end{align}
are  the respective Fourier-Wigner functions.
Here, $\bbE $ is the expectation with respect to the product
measure $\mu_\eps\otimes\bbP$. 
To simplify the notation we shall also write  $\widehat{
  W}_{\eps}(t,\eta,k)$ instead of $\widehat{
  W}_{\eps,+}(t,\eta,k).$

A straightforward calculation, using \eqref{basic:sde:2aas},
shows 
that  
\begin{align}
\label{012501-22}
\frac{d}{dt}\int_{\bbT}\bbE\big|\hat\psi^{(\eps)}(t,k)\big|^2dk
                 = {\frac{\ga}{\eps}\left(2 -\frac{1}{\mu}\right)
                 \left(T-\bbE[{\frak p}^{(\eps)}_0(t)]^2\right)}  
\end{align}
with  ${\frak p}^{(\eps)}_0(t):={\frak p}_0(t/\eps)$.  As a result we get
\begin{equation}
\label{02281a}
\eps\int_{\bbT}\bbE\big|\hat\psi^{(\eps)}(t,k)\big|^2dk
\le \eps\int_{\bbT}\bbE\big|\hat\psi^{(\eps)}(0,k)\big|^2 dk+
\left(2 -\frac{1}{\mu}\right) \ga T t,\quad t\ge0.
\end{equation}
Thus, 
we conclude from (\ref{02281a}) that (see~\cite{GMMP}) 
\begin{equation}\label{mar2302}
\sup_{t\in[0,\tau]}\|W^{(\eps)}(t)\|_{{\cal A}'}<\infty, \hbox{ for each $\tau>0$.}
\end{equation}
Hence $W^{(\eps)}(\cdot)$ is sequentially weak-$\star$ compact over   
$(L^1([0,\tau];{\cal A}))^\star$ for any $\tau>0$.

{The initial Wigner distribution
\begin{equation}\label{eq:20x}
 \widehat{ W}_\varepsilon(\eta,k) :=\  \widehat{ W}_\varepsilon(0,\eta,k),\quad (\eta,k)\in \bbT_{2/\eps}\times\bbT
\end{equation} 
is assumed to  converge $\star$-weakly, as $\eps\to0$, in ${\cal A}'$ to a non-negative function  $W_0\in L^1(\bbR\times \bbT)\cap C(\bbR\times \bbT)$.  
In addition, we suppose that 
there exist $C,\kappa>0$ such that 
\begin{equation}
\label{011812aa}
|\widehat W_\eps(\eta,k)|\le 
C\varphi(\eta),\quad (\eta,k)\in\bbT_{2/\eps}\times \bbT, \,\eps\in(0,1],
\end{equation} 
where
\begin{equation}
\label{011812c}
\varphi(\eta):=\frac{1}{(1+\eta^2)^{3/2+\kappa}}.
\end{equation} }

Define the Fourier-Laplace-Wigner functions
\begin{align}
\label{hwigner}
&\widehat{
  w}_{\pm,\varepsilon}(\la,\eta,k)=\eps\int_0^{+\infty}e^{-\la \eps t}\widehat{
  W}_{\pm,
\eps}(t,\eta,k)dt, \\
&
\widehat{
  y}_{\pm,\varepsilon}(\la,\eta,k)=\eps\int_0^{+\infty}e^{-\la \eps t}\widehat{
  Y}_{\pm,
\eps}(t,\eta,k)dt,\notag
\end{align}
where ${\rm Re}\,\la>0$, $(\eta,k)\in\bbT_{2/\eps}\times\bbT$. We
shall also write $\widehat{
  w}_{\varepsilon}(\la,\eta,k)$ instead of $\widehat{
  w}_{+,\varepsilon}(\la,\eta,k)$.

\subsection{Some additional notation}

Define
\begin{equation}
  \label{eq:bessel0J}
  J(t) = \int_{\bbT}\cos\left(\omega(k) t\right) dk,\quad t\in\bbR.
\end{equation}
Its Laplace transform
\begin{equation}
  \label{eq:2}
\tilde J(\la):=\int_0^{\infty}e^{-\la t}J(t)dt= \int_{\bbT}  \frac{\lambda}{\lambda^2 + \omega^2(k)} dk,\quad {\rm Re}\,\la>0.
\end{equation} 
\blue{One can easily see that
\begin{equation}
\label{012401-22}
|\tilde J(\la)|<\frac{1}{{\rm Re}\,\la}\quad\mbox{ for  ${\rm Re}\,\la>0$.}
\end{equation}}
Let
\begin{equation}
\label{tg}
\tilde g(\lambda) := ( 1 + \gamma \tilde J(\lambda))^{-1}.
\end{equation} 
{We have ${\rm Re}\,\tilde J(\la)>0$ for $\la\in \mathbb
  C_+:=[\la\in \mathbb C:\, {\rm Re}\,\la>0]$, thus in consequence
\begin{equation}
\label{012410}
|\tilde g(\lambda)|\le 1,\quad \la\in \mathbb C_+.
\end{equation} }
In addition, we have
 \begin{equation}
\label{tgJ}
(\tilde g\tilde J)(\lambda) =\frac{1}{\ga} ( 1-  \tilde g(\lambda))
{ = \frac{\tilde J(\lambda)}{1+\gamma \tilde J(\lambda)}
=\sum_{n=1}^{+\infty}(-\ga)^{n-1}\tilde{J}(\lambda)^n}.
\end{equation} 
\blue{The first two equalities in \eqref{tgJ} hold for all $\la\in \mathbb C_+$, while the
last one for ${\rm Re}\,\la>\ga$ (cf \eqref{012401-22}).}

{Since $|J(t)|\le 1$  we have  $|J^{\star,n}(t)|\le t^{n-1}/(n-1)!$, as the
  $n$-th convolution power involves the integration over
an $n-1$-dimensional simplex of size $t$.}
Therefore the series 
\begin{equation}
\label{011506-20s}
g_*(t):=\sum_{n=1}^{+\infty} (-\ga)^nJ^{\star,n}(t)
\end{equation}
defines a $C^\infty$ class function on $[0,+\infty)$ that satisfies
the following growth condition:
 there exists $C>0$ such that
$|g_*(t)|\le Ce^{\ga t}$, $t>0$. \blue{In addition, comparing the Laplace
transform of $g_*(t)$ with $( 1-  \tilde g(\lambda))/\ga$, as
expressed by the
utmost right hand side of \eqref{tgJ}, we conclude that
\begin{equation}
\label{gs}
\tilde g_*(\la)=\tilde g(\la)-1=-\ga (\tilde g\tilde J)(\lambda),\quad {\rm Re}\,\la>\ga.
\end{equation}}
Therefore $\tilde g(\la)$, given by \eqref{tg}, is the Laplace transform of the signed
measure $g(dt):=\delta_0(dt)+g_*(t)dt$.
Combining \eqref{tg}, \eqref{011506-20s} and \eqref{gs} we obtain
\begin{equation}
\label{011506-20}
\ga J\star
g(t)=\sum_{n=1}^{+\infty}(-1)^{n-1}\ga^nJ^{\star,n}(t)=-g_*(t),\quad t\ge0.
\end{equation}
\blue{It turns out, see Lemma \ref{lm011811-20} below, that $ J\star
g\in
L^2(\bbR)$ and ${\rm supp}\, J\star
g\subset [0,+\infty)$. This allows us to conclude the existence of
$\tilde g_*$ - the 
Laplace transform  of 
$g_*(\cdot)$ - and  equality \eqref{gs} for
all  $\la\in\mathbb C_+$.}


\subsection{Functions ${\widetilde{ g}}$ and ${\widetilde{ J}}$}

{Since the function $\tilde g(\cdot)$ is analytic on $ \mathbb C_+$
 we conclude, 
by the Fatou theorem, see e.g. p. 107 of \cite{koosis},
  that
\begin{equation}
\label{nu}
\tilde g(i\beta):=\lim_{\eps\to+0}\tilde g(\eps+i\beta) ,\quad\beta\in\bbR
\end{equation}
exists a.e.} In Section \ref{sec10.1} we show the following.
\begin{lemma}
\label{lm011811-20}
The holomorphic function $\tilde J \tilde g$ belongs to the Hardy
space $H^p(\mathbb C_+)$ for any $p\in(1,+\infty)$.
The limit
\begin{equation}
\label{030512-20}
(\tilde J \tilde g)(i\beta):=\lim_{\eps\to+0}(\tilde J \tilde g)(\eps+i\beta),\quad\beta\in\bbR
\end{equation}
exists both a.e. and in the $L^p(\bbR)$ sense for
$p\in(1,+\infty)$. 

In addition, there exists
\begin{equation}
\label{nuk}
\nu(k):=\lim_{\eps\to+0}\tilde g\big(\eps+i\om(k)\big) ,\quad k\in \Om_*,
\end{equation}
where $
\Om_*:=[k\in\bbT:\,\om'(k)=0,\quad\mbox{or}\quad \om(k)=0].
$
The function is continuous on $\bbT\setminus\Om_*$.
Moreover, for any $\delta>0$
there exists $C>0$ such that
\begin{equation}
\label{060512-20}
\Big|\tilde g\big(\eps+i\om(k)\big)
-\nu(k)\Big|\le C\eps,\quad  {\rm dist}\,\big(k, \Om_* \big)\ge\delta.
\end{equation}
\end{lemma}


To state our main result we need some additional notation.
Define the group velocity
\[
\bar\om'(k):=\om'(k)/(2\pi)
\]
and
\begin{equation}
\label{033110}
\wp(k):=\frac{\ga \nu(k)}{2|\bar\om'(k)|},\quad \fgeeszett(k):=\frac{\ga|\nu(k)|^2}{|\bar\om'(k)|},\quad p_+(k):=\left|1-\wp(k)\right|^2 ,\quad 
p_-(k):=|\wp(k)|^2  .
\end{equation}
It has been shown  in Section 10 of \cite{kors} that
\begin{equation}
\label{feb1402}
{\rm Re}\,\nu(k)=\left(1+\frac{\ga}{2|\bar\om'(k)|}\right)|\nu(k)|^2
\end{equation}
and 
\begin{equation}\label{mar1528}
p_+(k)+p_-(k)=1- \fgeeszett(k)
\le 1 ,
\end{equation}
so that, in particular, we have
\begin{equation}\label{feb1420}
0\le \fgeeszett(k)\le 1,\quad k\in\bbT.
\end{equation}
In the model considered in \cite{kors}  the coefficients $p_+(k)$, $p_-(k)$ and $\fgeeszett(k)$ have expressed, see \cite[Theorem 2.1]{kors},  the probabilities of
a phonon being transmitted, reflected and absorbed at the interface
$[x=0]$.

In our present situation the absorption probability needs to be  modified. In
addition, the phonon  can be also scattered at the interface with
outgoing frequency $\ell$ with some scattering rate  $r(k,\ell)$. To
be more precise we introduce the following notation
\begin{equation}
\label{p-sc-prod}
 p_{\rm abs}:=\frac{
   1}{1-\Gamma/\mu}\left(1-\frac{1}{2\mu}\right),\quad p_{\rm sc}(\ell):=
 \frac{1}{2\mu(1-\Gamma/\mu)}
|\nu(\ell)|^2,
\end{equation}
where
\begin{equation}
\label{Gamma}
\Gamma:=\frac{\ga}{2\pi}\int_{\bbR}|\tilde J\tilde
  g(i\beta')|^2  d\beta'.
\end{equation}

The following result holds.
\begin{lemma}
\label{lm013011-20}
{For any $\ga>0$
we have
\begin{equation}
\label{022712-20}
\Gamma+\frac{1}{2}\int_{\bbT}|\nu(\ell)|^2 d\ell=\frac{1}{2}.
\end{equation}}
In addition, if 
 $\mu\ge1/2$, then
\begin{equation}
\label{fund-ident2}
p_{\rm abs}+ \int_{\bbT}p_{\rm sc}(\ell)d\ell=1.
\end{equation}
\end{lemma}
The proof of the lemma is contained in Section \ref{sec10.2}.

{ \begin{remark}{\em It turns out, see \cite[Theorem 4. part
      iii)]{KO20}, that for any unimodal dispersion relation we have
 $|\nu(\ell)|>0$, except possibly $\ell=0$, or $1/2$.
 Thanks to the identity  \eqref{022712-20} below, we have then
\begin{equation}
\label{Gamma-a}
\Gamma<\frac{1}{2}\le \mu.
\end{equation}
Therefore, in particular, the coefficients defined in
\eqref{p-sc-prod} are strictly positive for $\mu>1/2$ and $\ell\not\in\{0,1/2\}$.
}
\end{remark}}

\subsection{The main result} 

{{For brevity sake, we use the notation }
\begin{equation*}
[[0, a]] := \begin{cases}
[0, a], &\text{if $a>0$}\\
[a, 0], &\text{if $a<0$.}
\end{cases}
\end{equation*}
{The main result of the paper can be formulated as follows. }
\begin{thm}
\label{main:thm}
Suppose that the initial conditions and the dispersion relation
satisfy the above assumptions. 
Then, for any $\tau>0$ and $G\in L^1\left([0,\tau];{\cal A}\right)$ we have 
\begin{equation}
\label{022410}
\lim_{\eps\to0}\int_0^\tau\langle G(t),W_\eps(t)\rangle dt=\int_0^\tau
dt\int_{\bbR\times\bbT}G^*(t,x,k)W(t,x,k)dxdk,
\end{equation}
where
\begin{equation}\label{010304}
  \begin{split}
&W\left(t,x,k\right)
=   
W_0\left(x-\bar{\om}'(k)t,k\right) 1_{[[0,\bar{\om}'(k)t]]^c}(x) \vphantom{\int_0^1}
+p_+(k)W_0\left(x-\bar{\om}'(k)t,k\right)1_{[[0,\bar{\om}'(k)t]]}(x)
\\
&
+p_-(k) W_0\left(-x+\bar\om'(k) t,-k\right)1_{[[0,\bar \om'(k)t]]}(x)
\\
&
 +\fgeeszett (k) 1_{[[0,\bar{\om}'(k)t]]}(x) \int_{\bbT}
W_0\left({  \frac{\bar{\om}'(\ell)}{\bar{\om}'(k)}\big(x-\bar{\om}'(k)t\big)},\ell\right)
p_{\rm sc}(\ell)d\ell 
\\
&
+
  p_{\rm abs} \, \fgeeszett (k) T1_{[[0,\bar{\om}'(k)t]]}(x). 
\end{split}
\end{equation}
\end{thm}
The proof of this result is given in Section \ref{sec:nul-init-cond}.

{ The limit dynamics can be characterized as follows:} $W(t,x,k)$ describes the energy density in
$(x,k)$ at time $t$ of the phonons initially distributed according to
$W_0(x,k)$.
The first term corresponds then to the ballistic transport of those phonons which did not cross $\{x=0\}$ up to time $t$. 
The second and third   terms correspond, respectively, 
to the transmission and reflection of the phonons at the boundary
point $\{x=0\}$ with probabilities  $p_+(k)$ and $p_-(k)$, respectively.
The fourth term describes the phonon scattering
that occurs at the interface. The phonon with frequency $\ell$, arriving
at the interface with the velocity $\bar\om'(\ell)$ is scattered with frequency $k$ at the rate
$\fgeeszett(\ell) p_{\rm sc}(k)$ and moves away from the interface
with the velocity  $\bar\om'(k)$.
Finally, the last term {in the right side} of \eqref{010304} describes the
$k$-phonon production of the thermostat at the rate $p_{\rm abs} \fgeeszett(k)T$.
From \eqref{mar1528} and \eqref{fund-ident2} we conclude that
\begin{equation}
\label{balance}
1 - p_+(\ell) - p_-(\ell) -\fgeeszett(\ell)\int_{\bbT} p_{\rm sc}(k)dk=
p_{\rm abs} \fgeeszett(\ell),\quad
\ell\in\bbT.
\end{equation}
Therefore, the $\ell$-phonon is absorbed by the thermostat with
probability $p_{\rm abs} \fgeeszett(\ell)$.
\blue{Note that in the special case when the thermostat operates by
  the flip of the
  momentum, which happens  when $\mu=1/2$, there is no absorption, as according to
  \eqref{p-sc-prod} we have $p_{\rm abs} =0$. This is consistent with
  the fact that the total energy of the chain is then conserved, see \eqref{012501-22}.}

Our result can be written as a boundary value problem.
Note that
$W(t,x,k)$ solves the homogeneous transport equation
\begin{equation}\label{feb1406}
\partial_tW(t,x,k)+ \bar\om'(k) \partial_x W(t,x,k)=0,
\end{equation}
away from the boundary point $\{x=0\}$. 

Let
\[
W(t,0^\pm,k):=\lim_{x\to\pm0}W(t,x,k).
\] 
If $k\in\bbT_+$ ($k>0$), then  
\begin{align}
\label{010304zzy}
&W\left(t,0^+,k\right)
=  p_+(k) W\left(t,0^-,k\right)
+p_-(k) W\left(t,0^+,-k\right)
+
  p_{\rm abs} \, \fgeeszett(k) T
\\
&
+\fgeeszett(k)  \int_{\bbT_+}
 W\left(t,0^-,\ell\right)
  p_{\rm sc}(\ell)d\ell +  \fgeeszett(k)   \int_{\bbT_+}
 W\left(t,0^+,-\ell\right)
  p_{\rm sc}(\ell)d\ell.\nonumber
\end{align}
If, on the other hand, $k\in\bbT_-$ ($k<0$), then  
\begin{equation}\label{feb1408}
\begin{split}
&W\left(t,0^-,k\right)
=  p_+(k) W\left(t,0^+,k\right)
+p_-(k) W\left(t,0^-,-k\right)
+
  p_{\rm abs} \, \fgeeszett(k)  T
\\
&
+ \fgeeszett(k)   \int_{\bbT_-}
 W\left(t,0^+,\ell\right)
  p_{\rm sc}(\ell)d\ell + \fgeeszett(k)    \int_{\bbT_-}
 W\left(t,0^-,-\ell\right)
  p_{\rm sc}(\ell)d\ell.\nonumber
\end{split}
\end{equation}

\section{The solution of \eqref{basic:sde:2aas} and  its Laplace-Fourier-Wigner distribution}
\label{sec:wave-fun}

In this section, we obtain an explicit expression for  the solution of
the wave function \eqref{basic:sde:2aas}. 
The mild formulation of  the equation  reads as follows
\begin{equation}
\begin{split}
 \label{mild}
 \hat\psi(t,k) &=e^{-i\om(k)t}\hat\psi(k)-\frac{i}{\mu}\int_0^t
 e^{-i\om(k)(t-s)}{\frak p}_0(s-) dN(\ga\mu s)\\
&
+i \int_0^t
 e^{-i\om(k)(t-s)}\tilde p(s-)dN(\ga\mu s),
\end{split}
 \end{equation}
where ${\tilde
   p}(t)$ is given by \eqref{tp}. Letting
 \begin{align}
\label{eq:bessel0}
 {\frak p}_0^0(t) : ={\rm Im} \left(\int_{\bbT}e^{-i\om(k)t}\hat\psi(k) dk\right)
\end{align}
  we conclude the following closed equation on the momentum at $y=0$:
\begin{equation}
\begin{split}
 \label{momentum}
 {\frak p}_0(t) &={\frak p}_0^0(t) -\frac{1}{\mu}\int_0^tJ(t-s){\frak p}_0(s-) dN(
 \ga\mu s)+ \int_0^tJ(t-s)\tilde p(s-)dN( \ga\mu s).
\end{split}
 \end{equation}
Equation \eqref{mild} is linear, so  its solution  can
be written as the sum of the solution $\hat\psi_1(t,k)$  corresponding
to the null initial data $\hat\psi(k)\equiv 0$ and the solution $\hat\psi_2(t,k)$ of the
homogeneous equation corresponding to $\tilde p(t)\equiv0$.

More precisely, suppose that $\hat\psi_1(t,k)$  is the
solution of 
\begin{equation}
\begin{split}
 \label{basic:sde:2aa-n}
 d\hat\psi_1(t,k) &= -i\om(k) \hat\psi_1(t,k)
dt
 +i\left({\tilde
   p}(t -)-\frac{1}{\mu}{\frak p}_{0,1}(t-)\right)dN(\ga\mu t),
 \\
\hat\psi_1(0,k) &\equiv 0
\end{split}
 \end{equation} 
and   $\hat \psi_2(t,k)$ satifsies
\begin{equation}
\begin{split}
 \label{basic:sde:3x-n}
 d\hat\psi_2(t,k) &= -i\om(k)\hat\psi_2 (t,k)
-\frac{i}{\mu}{\frak p}_{0,2}  (t-)dN(\ga\mu t ),
 \\
\hat\psi_2 (0,k) & = \hat\psi(k).
\end{split}
\end{equation}
Here
$$
{\frak p}_{0,j} (t):={\rm Im}
\int_{\bbT}\hat\psi_j(t,k) dk,\quad j=1,2.
$$ 
Then
\begin{equation}
\label{psi1-2}
\hat\psi(t,k)=\hat\psi_1(t,k)+\hat\psi_2(t,k).
\end{equation}
The respective Fourier-Wigner functions are defined as 
$$
\widehat{ W}_\varepsilon^{j_1,j_2}(t,\eta,k) :=\frac{ \varepsilon}{2}
\bbE  \left[\hat\psi^*_{j_1}\left( \frac{t}{\eps} , k- \frac{\varepsilon\eta}2\right) 
  \hat\psi_{j_2} \left( \frac{t}{\eps} , k +  \frac{\varepsilon\eta}2\right)
\right],\quad j_1,\,j_2\in\{1,2\}.
$$ 
Since the process $\big(\tilde p(t)\big)_{t\ge0}$ is independent of
the initial data field $\big(\hat\psi(k)\big)_{k\in\bbT}$ we conclude
easily that
$$
\widehat{ W}_\varepsilon^{j_1,j_2}(t,\eta,k) \equiv 0,\quad\mbox{if
}j_1\not= j_2.
$$ 
Therefore,
\begin{equation}
\label{W12}
\widehat{ W}_\varepsilon(t,\eta,k) :=\frac{ \varepsilon}{2}
\bbE  \left[\hat\psi^*\left( \frac{t}{\eps} , k- \frac{\varepsilon\eta}2\right) 
  \hat\psi\left( \frac{t}{\eps} , k +  \frac{\varepsilon\eta}2\right)
\right]=\widehat{ W}^{1,1}_\varepsilon(t,\eta,k)+\widehat{ W}^{2,2}_\varepsilon(t,\eta,k).
\end{equation}
Accordingly, the respective Laplace-Fourier-Wigner transforms satisfy
\begin{align}
\label{hwigner-12}
\widehat{
  w}_\varepsilon(\la,\eta,k)=\widehat{
  w}_\varepsilon^{1,1}(\la,\eta,k)+\widehat{
  w}_\varepsilon^{2,2}(\la,\eta,k),
\end{align}
where 
$$
\widehat{
  w}_\varepsilon(\la,\eta,k)=\int_0^{+\infty}e^{-\la t}\widehat{
  W}_\varepsilon(t,\eta,k)dt,\quad (\eta,k)\in\bbT_{2/\eps}\times\bbT
$$
and ${\rm Re}\,\la>0$. The definitions of $\widehat{
  w}_\varepsilon^{j,j}$, corresponding to $\widehat{ W}_\varepsilon^{j,j}(t,\eta,k)$, $j=1,2$ are analogous.


\subsection{Solving \eqref{basic:sde:2aas} for the null initial data}

\label{sec4}

We suppose that $\hat\psi(0,k)\equiv0$.
Let $s_0:=t$, $\Delta_1(t):=[0,t]$ and
 \begin{align*}
 \Delta_n(t):=[(s_1,\ldots,s_n):\,t> s_1>s_2> \ldots> s_n>0],\quad n\ge2.
\end{align*}
Iterating \eqref{momentum} and remembering that ${\frak p}_0^0(t)\equiv0$
we can write
\begin{equation}
 \label{momentum1}
 {\frak p}_{0,1}(t) =
 \sum_{n=1}^{+\infty}\left(-\frac{1}{\mu}\right)^{n-1}\int_{\Delta_n(t)}\prod_{j=1}^{n}J(s_{j-1}-s_j)\tilde
 p(s_n-)dN( \ga\mu s_1)\ldots dN( \ga\mu s_n),
\end{equation}
with $s_0:=t$. Therefore, substituting for the momentum into the
respective form of \eqref{mild} we get
\begin{equation}
 \label{mild-b10}
\hat\psi_1(t,k) =i\int_0^t
 e^{-i\om(k)(t-s)}\left(\tilde p(s)-\frac{1}{\mu}{\frak p}_0(s-) \right)dN(\ga\mu
 s)=\sum_{n=1}^{+\infty}\hat\psi_{1,n}(t,k), 
\end{equation}
where
\begin{equation}
\begin{split}
 \label{mild-b1}
&
\hat\psi_{1,1}(t,k):=i\int_0^t
 e^{-i\om(k)(t-s)}\tilde p(s-) dN( \ga\mu s),\\
&
\hat\psi_{1,n}(t,k):=\left(-\frac{1}{\mu}\right)^{n-1} i\int_{\Delta_{n}(t)}
 e^{-i\om(k)(t-s_1)}\\
&
\qquad \qquad \qquad\times\prod_{j=1}^{n-1}J(s_{j}-s_{j+1})\tilde
 p(s_{n}-)dN( \ga\mu s_1)\ldots dN( \ga\mu s_{n}),\quad n\ge 2.
\end{split}
 \end{equation}

\subsection{The case $T=0$ and non-zero initial data}

\label{sec5}

The mild formulation of \eqref{basic:sde:3x-n} is as follows
\begin{equation}
\begin{split}
 \label{mild-3}
 \hat\psi_2(t,k) &=e^{-i\om(k)t}\hat\psi(k)-\frac{i}{\mu}\int_0^t
 e^{-i\om(k)(t-s)}{\frak p}_{0,2}(s-) dN(\ga\mu s).
\end{split}
 \end{equation}
From here we conclude the following closed equation on the momentum at $y=0$:
\begin{equation}
\begin{split}
 \label{momentum-3z}
 {\frak p}_{0,2}(t) &={\frak p}_0^0(t) -\frac{1}{\mu}\int_0^tJ(t-s){\frak p}_{0,2}(s-) dN( {\ga\mu}s),
\end{split}
 \end{equation}
where ${\frak p}_0^0(t)$ is given by \eqref{eq:bessel0}.
The solution of \eqref{momentum-3z} is given by 
\begin{equation}
\begin{split}
 \label{momentum-3a}
& {\frak p}_{0,2}(t) ={\frak p}_0^0(t)
 +\sum_{n=1}^{+\infty}\left(-\frac{1}{\mu}\right)^n\int_{\Delta_n(t)}J(t-s_1)\ldots
 J(s_{n-1}-s_n)\\
&
\qquad \qquad \qquad \qquad \times{\frak p}_0^0(s_n) dN( \ga\mu s_1)\ldots  dN( \ga\mu s_n).
\end{split}
 \end{equation}
Substituting into \eqref{mild-3} we get 
\begin{equation}
  \label{eq:sol-xx0}
\hat\psi_2(t,k)=\sum_{n=0}^{+\infty}
  \hat\psi_{2,n}(t,k),
 \end{equation}
where
\begin{equation}
  \label{eq:sol-xx}
  \begin{split}
&
\hat\psi_{2,1}(t,k):=-\frac{i}{\mu}\int_0^t
 e^{-i\om(k)(t-s)}{\frak p}_0^0(s) dN(\ga\mu s)\\
&
\hat\psi_{2,n}(t,k):=-i
\sum_{n=1}^{+\infty}\left(-\frac{1}{\mu}\right)^{n+1}
\int_{\Delta_n(t)} e^{-i\om(k)(t-s_1)}\prod_{j=1}^{n-1}
J(s_j-s_{j+1})\\
&
\qquad \qquad \qquad \qquad \times{\frak p}_0^0(s_n) dN( \ga\mu s_1)\ldots  dN( \ga\mu
 s_n),\quad n\ge2.
  \end{split}
\end{equation}

\section{The limit in case of null initial data - the phonon creation term}
\label{sec:phonon-creation-term}

Consider first the case when the null initial data,
i.e. $\hat\psi_2(t,k)\equiv0$.  
Then, 
\begin{equation}
\label{010112-20}
\widehat{ w}_{\eps} (\la,\eta,k) =\widehat{ w}_{\eps}^{(1,1)}
(\la,\eta,k) .
\end{equation}
We wish to use the chaos expansion,
corresponding to the  Poisson process $(N(t))_{t\ge0}$ to represent
the Laplace-Fourier-Wigner  function $\widehat{ w}_{\eps} (\la,\eta,k)
$.    
\begin{lemma}
\label{lm010112-20}
Suppose that $\mu> 1/2$. The following formula holds
\begin{equation}
\label{021601-19-1}
\widehat{ w}_{\eps} (\la,\eta,k) =\frac{ \eps T\ga }{\la}\left(1-\frac{1}{2\mu}\right)
   \int_0^{+\infty}e^{-\la \eps s}
\bbE\left[\hat\chi^*\left(s,k-\frac{\eps\eta}{2}\right) \hat\chi\left(s,k+\frac{\eps\eta}{2}\right)\right]ds
\end{equation}
for any $\la\in \mathbb C_+$, $(\eta,k)\in\bbT_{2/\eps}\times\bbT$ and
$\eps>0$.
Here
\begin{align}
\label{hatc}
&\hat\chi(t,k):=\exp\left\{-i\om\left(k\right)t\right\}\\
&
+\sum_{n=1}^{+\infty}\left(-\frac{1}{\mu}\right)^{n}\int_{\Delta_{n}(t)}\exp\left\{-i\om\left(k\right)(t-s_1)\right\}\prod_{j=1}^{n}J(s_{j}-s_{j+1}) dN( \ga\mu s_1)\ldots dN( \ga\mu s_{n}), \notag
\end{align}
with $s_{n+1}:=0$.

\blue{ If, on the other hand $\mu= 1/2$ and
 $\eps,\ga>0$, then
\begin{equation}
\label{021601-19aa}
 \widehat{ w}_{\eps} (\la,\eta,k)= 0,\quad 
 \la\in\mathbb C_+, \,(\eta,k)\in\bbR\times\bbT.
\end{equation}}
\end{lemma}
\blue{\begin{remark}
Note that \eqref{021601-19aa} is consistent with the physical
interpretation of the model. Namely, we have assumed that  
initially the energy of the chain is   null. On the other hand the
momentum flip mechanism of the thermostat, that corresponds to the case
$\mu=1/2$,   conserves the total energy
of the system.
\end{remark}}

\subsubsection*{Proof of Lemma \ref{lm010112-20}}
\blue{The series appearing on the right hand side of \eqref{hatc}
  converges in the $L^1$ sense. Indeed,   since $|J(t)|\le 1$ its
  terms are dominated by the respective terms of the series
\begin{equation}
\label{Theta}
\Theta(t):=1
+\sum_{n=1}^{+\infty}\left(\frac{1}{\mu}\right)^{n}\int_{\Delta_{n}(t)}dN( \ga\mu s_1)\ldots dN( \ga\mu s_{n}).
\end{equation}
The process $\Theta(t)$ is the unique solution of  the stochastic differential equation
$d\Theta(t)=\big(\Theta(t-)/\mu\big) dN( \ga\mu t)$, $\theta(0)=1$ and
is given by the stochastic exponential, see
e.g. \cite[Theorem II.8.37, p. 84]{protter},
$$
\Theta(t)=\exp\left\{N( \ga\mu t)\log\left(1+\frac{1}{\mu}\right)\right\}.
$$
We have  $|\hat\chi(t,k)|\le \Theta(t)$, therefore
$$
\bbE|\hat\chi(t,k)|^2\le \bbE\Theta^2(t)=\exp\left\{\ga\mu t\Big[\exp\left\{2\log\left(1+\frac{1}{\mu}\right)\right\}-1\Big]\right\}
$$
and the right hand side of  \eqref{021601-19-1} is well defined, as an
element of ${\cal A}'$ (see \eqref{norm-ta01}), at
least for
${\rm Re}\,\la>2\ga \eps^{-1}$. In what follows we show that equality \eqref{021601-19-1} 
holds for this range of 
$\la$. Note that this implies the validity of \eqref{021601-19-1}  for
all $\la\in\mathbb C_+$. Indeed, if $\mu=1/2$, then by the analytic continuation we conclude that $\widehat{ w}_{\eps} (\la,\eta,k)=0$
for all ${\rm Re}\,\la>0$ and the formula \eqref{021601-19aa} follows.}

\blue{For $\mu>1/2$, the equality of the Laplace transforms, see
  \eqref{021601-19-1}, for ${\rm Re}\,\la>2\ga \eps^{-1}$  implies in particular, when $\eta=0$,
that 
$$
   T\ga
   \left(1-\frac{1}{2\mu}\right)\bbE\left|\hat\chi\left(\frac{t}{\eps},k\right)\right|^2=\frac{\eps}{2}\frac{d}{dt}\bbE\left|\hat\psi_1\left(\frac{t}{\eps},k\right)\right|^2,\quad t\ge0.
$$
In light of \eqref{012501-22}, this allows us to extend the validity
of  \eqref{021601-19-1} to all ${\rm Re}\,\la>0$.}

\blue{Now we proceed with the proof of \eqref{021601-19-1} for ${\rm Re}\,\la>2\ga \eps^{-1}$.} Substituting from \eqref{mild-b1} we get
\begin{equation}
\label{021601-19}
\widehat{ w}_{\eps} (\la,\eta,k) =\sum_{n,m=1}^{+\infty}\widehat{
  w}_{\eps,n,m} (\la,\eta,k), 
\end{equation}
where
\begin{align*}
\widehat{ w}_{\eps,n,m} (\la,\eta,k):=
\frac{ \varepsilon}{2} \int_0^{+\infty}e^{-\eps\la t}\bbE\left[\hat\psi^\star_{1,n}\left( t, k- \frac{\varepsilon\eta}2\right) 
  \hat\psi_{1,m}\left( t , k +  \frac{\varepsilon\eta}2\right)\right]dt
\quad n,m\ge 1.\notag
\end{align*}
\blue{The convergence of the series follows by the
comparison with the series defining the stochastic exponential, see \eqref{Theta}.}
Note that for $s>s'$ 
{ 
$$
\bbE\Big[\tilde p(s-) \tilde p(s'-),\, N( \ga\mu  s-)-N( \ga\mu  s') \ge 1\Big]=0.
$$
The above implies that}
\begin{align*}
&
\widehat{ w}_{\eps,n,m} (\la,\eta,k)=
\frac{ \eps^2}{2}\left(-\frac{1}{\mu}\right)^{n+m}
   \int_0^{+\infty}e^{-\la \eps t}dt \\
&
\times\bbE\left[\int_{\Delta_{n}(t)}dN( \ga\mu s_1)\ldots dN( \ga\mu  s_{n}) \int_{\Delta_{m}(t)}dN( \ga\mu  s_1')\ldots dN( \ga\mu  s_{m}') \exp\left\{i\om\left(k-\frac{\eps\eta}{2}\right)(t-s_1)\right\}\right.\\
&
 \times\left.
\exp\left\{-i\om\left(k+\frac{\eps\eta}{2}\right)(t-s_1')\right\}\prod_{j=1}^{n-1}J(s_{j}-s_{j+1}) \prod_{j=1}^{m-1}J(s_{j}'-s_{j+1}')\tilde
 p(s_{n}-) \tilde
 p(s_{m}'-)\right]
\end{align*}
\begin{align*}
&
=
\eps^2T\ga  \left(1-\frac{1}{2\mu}\right) \left(-\frac{1}{\mu}\right)^{n+m}
   \int_0^{+\infty}e^{-\la \eps t}dt \int_0^tds \\
&
\times \bbE\left[\int_{\Delta_{n-1}(t-s)}dN( \ga\mu s_1)\ldots dN( \ga\mu s_{n-1}) \int_{\Delta_{m-1}(t-s)}dN( \ga\mu s_1')\ldots dN( \ga\mu s_{m-1}')\right.\\
&
 \times\left.
 \exp\left\{i\om\left(k-\frac{\eps\eta}{2}\right)(t-s-s_1)\right\}\exp\left\{-i\om\left(k+\frac{\eps\eta}{2}\right)(t-s-s_1')\right\}\prod_{j=1}^{n-1}J(s_{j}-s_{j+1}) \prod_{j=1}^{m-1}J(s_{j}'-s_{j+1}')\right].
\end{align*}
Here $s_n=s_m':=0$.
Integrating out the $t$ variable we get
\begin{align*}
&
\widehat{ w}_{\eps,n,m} (\la,\eta,k)=
\frac{ \eps \ga T}{\la}\left(1-\frac{1}{2\mu}\right) \left(-\frac{1}{\mu}\right)^{n+m}
   \int_0^{+\infty}e^{-\la \eps s}
  \exp\left\{i\left[\om\left(k-\frac{\eps\eta}{2}\right)-\om\left(k+\frac{\eps\eta}{2}\right)\right]s\right\}ds\\
&
\times
\bbE\left[\int_{\Delta_{n-1}(s)}dN( \ga\mu s_1)\ldots dN( \ga\mu s_{n-1}) \int_{\Delta_{m-1}(s)}dN( \ga\mu s_1')\ldots dN( \ga\mu s_{m-1}')\right.\\
&
 \times\left.
 \exp\left\{-i\om\left(k-\frac{\eps\eta}{2}\right)s_1\right\}\exp\left\{i\om\left(k+\frac{\eps\eta}{2}\right)s_1'\right\}\prod_{j=1}^{n-1}J(s_{j}-s_{j+1}) \prod_{j=1}^{m-1}J(s_{j}'-s_{j+1}')\right]
\end{align*} 
for $n,m\ge1$. Summing out over $n,m$ we conclude
\eqref{021601-19-1}. 
\qed

\bigskip

Next, we write the Poisson chaos decomposition
of the random field $\hat\chi(t,k)$. Let
\begin{equation}
\label{phi-t1}
\phi(t,k):=\int_0^te^{-i\om(k)(t-s)}g(ds).
\end{equation}
Define, the cadlag martingale
\begin{equation}
\label{tN}
\tilde N( t):=N(t)-t,\quad t\ge0.
\end{equation}
\begin{lemma}
\label{lm020112-20}
The following expansion holds
\begin{equation}
\label{mild-1}
\hat\chi(t,k)=\sum_{n=0}^{+\infty}\hat\chi_n(t,k),
\end{equation}
where
\begin{align}
\label{021709-20}
&
\hat\chi_0(t,k):= \phi(t,k), \notag\\
&
\hat\chi_n(t,k) :=\left(-\frac{1}{\mu}\right)^{n}\int_{\Delta_n(t)}\phi(t-s_1,k)\prod_{j=1}^{n}J\star
  g(s_{j}-s_{j+1})\\
&
\qquad  \qquad  \qquad \times d\tilde N( \ga\mu s_1)\ldots d\tilde N(
  \ga\mu s_{n}),\quad n\ge1.\notag
\end{align}
\end{lemma}
\proof
Writing $N( \ga\mu t)=\tilde N( \ga\mu t)+\ga\mu t$, where $\left(\tilde
  N( \ga\mu t)\right)_{t\ge0}$ is a cadlag martingale we obtain
\begin{equation}
\begin{split}
 \label{momentum1a}
 &
\hat\chi(t,k)= \exp\left\{-i\om\left(k\right)t\right\}+\sum_{n=1}^{+\infty}(-\ga)^{n}\int_{\Delta_{n}(t)}\exp\left\{-i\om\left(k\right)(t-s_1)\right\}\prod_{j=1}^{n}J(s_{j}-s_{j+1}) ds_1\ldots ds_{n}
 \\
 &
 +\sum_{n=1}^{+\infty}\left(-\frac{1}{\mu}\right)^{n}\sum_{k=1}^{n-1}(\ga\mu)^k\sum_{{\bf
     i}\in {\cal
     I}_k^n}\int_{\Delta_n(t)}\exp\left\{-i\om\left(k\right)(t-s_1)\right\}\prod_{j=1}^{n}J(s_{j}-s_{j+1}) d{\bf s}_{ {\bf i}} \prod_{j\not\in  {\bf i}}
 d\tilde N(\ga\mu s_j)
\\
 &+\sum_{n=1}^{+\infty}\left(-\frac{1}{\mu}\right)^{n}\int_{\Delta_n(t)}\exp\left\{-i\om\left(k\right)(t-s_1)\right\}\prod_{j=1}^{n}J(s_{j}-s_{j+1}) d\tilde N(\ga\mu s_1)\ldots d\tilde N(\ga\mu s_n).
\end{split}
 \end{equation}
For $1\le k\le n$ we denote by 
${\cal I}_k^n$ the set of all ordered $k$-indices ${\bf i}:\,1\le
i_1<\ldots<i_k\le n$. We shall also use the abbreviation
$
d{\bf s}_{{\bf i}} :=\prod_{j\in {\bf i}} ds_{j}.
$

Using \eqref{011506-20} we can combine the first two terms in the right hand side of
\eqref{momentum1a} and obtain that they are equal to $\phi(t,k) $ (cf \eqref{phi-t1})


Changing the order of summation in the remaining two expressions in the right hand side of \eqref{momentum1a} we conclude that their sum equals
\begin{align*}
&\sum_{n=1}^{+\infty}\left(-\frac{1}{\mu}\right)^{n}\sum_{r_1=0}^{+\infty}\sum_{r_2,\ldots,r_{n}=1}^{+\infty}\int_{\Delta_{n}(t)}\exp\left\{-i\om\left(k\right)(t-s_1)\right\}\\
&
\times\prod_{j=1}^{n}(-\ga)^{r_j-1} J^{\star,
  r_j}(s_{j}-s_{j+1})d\tilde N( \ga\mu s_1)\ldots d\tilde N(
  \ga\mu s_{n})
\end{align*}
Using formula \eqref{011506-20} the above expression can be rewritten in the form
\begin{align*}
&\sum_{n=1}^{+\infty}\left(-\frac{1}{\mu}\right)^{n}\int_{\Delta_n(t)}\left(\int_0^{t-s_1}\exp\left\{-i\om\left(k\right)(t-s_1-\si)\right\}g(d\si)\right)\\
&
\times \prod_{j=1}^{n}J\star
  g(s_{j}-s_{j+1})d\tilde N( \ga\mu s_1)\ldots d\tilde N(
  \ga\mu s_n)\\
&
=\sum_{n=1}^{+\infty}\left(-\frac{1}{\mu}\right)^{n}\int_{\Delta_n(t)}\phi(t-s_1,k)\prod_{j=1}^{n}J\star
  g(s_{j}-s_{j+1})d\tilde N( \ga\mu s_1)\ldots d\tilde N(
  \ga\mu s_{n})
\end{align*}
and \eqref{mild-1}, with \eqref{021709-20} follow.
\qed

\bigskip


Coming back to calculation of the asymptotics of $\widehat{ w}_{\eps}
(\la,\eta,k) $ given by \eqref{010112-20} we have the following
result.
\begin{prop}
\label{prop010112-20}
For any $\ga>0$ the parameter $\Gamma$, defined by \eqref{Gamma},
belongs to  $(0,1/2)$. In addition, for any
$\mu>1/2$, $\gamma>0$,
 $\la\in\mathbb C_+$ and $(\eta,k)\in\bbR\times\bbT$ we have
\begin{equation}
\label{021601-19zz}
\lim_{\eps\to0+}\widehat{ w}_{\eps} (\la,\eta,k)= \frac{  \ga T|\nu(k)|^2}{(1-\Gamma/\mu)\la(\la+i\om'(k)\eta)}\left(1-\frac{1}{2\mu}\right).
\end{equation}
\end{prop}
\proof
We can use the $L^2(\bbP)$
orthogonality of the terms of the expansion \eqref{mild-1}, with
\eqref{021709-20}. 
For ${\rm Re}\,\la>0$ sufficiently large  we get
\begin{equation}
\label{021601-19z}
\widehat{ w}_{\eps} (\la,\eta,k) =\sum_{n=0}^{+\infty}\widehat{
  w}_{\eps}^{(n)} (\la,\eta,k), 
\end{equation}
where
\begin{align}
\label{011709-20}
&\widehat{ w}_{\eps}^{(0)} (\la,\eta,k):=\frac{ \eps T\ga }{\la}\left(1-\frac{1}{2\mu}\right)\int_0^{+\infty}
e^{-\eps\la t}
   \phi^\star\left(t,k-\frac{\eps\eta}{2}\right) \phi\left(t,k+\frac{\eps\eta}{2}\right) dt  ,\notag\\
&
\widehat{ w}_{\eps}^{(n)} (\la,\eta,k):=\frac{ \eps T\ga }{\la}\left(1-\frac{1}{2\mu}\right)\left(\frac{\ga}{\mu}\right)^n\int_0^{+\infty}
e^{-\eps\la t}dt \int_{\Delta_n(t)}
  \phi^\star\left(t-s_1,k-\frac{\eps\eta}{2}\right)
  \phi\left(t-s_1,k+\frac{\eps\eta}{2}\right) \\
&
\times \prod_{j=1}^{n}\left(J\star
  g(s_{j}-s_{j+1}) \right)^2ds_1\ldots ds_n,\quad n\ge1\notag
\end{align}
\blue{In what follows, see \eqref{032501-22} below, we show that \eqref{021601-19z} in fact holds for all
$\la\in\mathbb C_+$.}

\subsection*{Computation of $\widehat{ w}_{\eps}^{(0)} (\la,\eta,k)
  $}

Thanks to \eqref{021709-20} and \eqref{011709-20} we have
\begin{equation}
\label{022601-22}
\widehat{ w}_{\eps}^{(0)} (\la,\eta,k)=\frac{ \eps T\ga }{\la}\left(1-\frac{1}{2\mu}\right)\int_0^{+\infty}\int_0^{+\infty}dt dt'
e^{-\eps\la
  (t+t')/2}\delta(t-t') \phi^\star\left(t,k-\frac{\eps\eta}{2}\right)
  \phi\left(t',k+\frac{\eps\eta}{2}\right)
  \end{equation}
Using
\begin{equation}
\label{delta}
\delta(t-t')=\frac{1}{2\pi}\int_{\bbR}e^{i\beta(t-t')}d\beta,
\end{equation}
we can write
\begin{align}
\label{012601-22}
&
\widehat{ w}_{\eps}^{(0)} (\la,\eta,k)=\frac{ \eps T\ga }{(2\pi)\la}\left(1-\frac{1}{2\mu}\right)\int_{\bbR}d\beta \int_0^{+\infty}e^{-(\eps\la/2 - i\beta) t} dt\int_0^{t}\exp\left\{i\om\left(k-\frac{\eps\eta}{2}\right)(t-s)\right\}g(ds)
\\
&
\times 
\int_0^{+\infty} e^{-(\eps\la/2+i\beta) t'} dt'\int_0^{t'}\exp\left\{-i\om\left(k+\frac{\eps\eta}{2}\right)(t'-s')\right\}g(ds').\notag
\end{align}
\blue{\begin{remark}
The use of formula \eqref{delta} in derivation of \eqref{012601-22} is a bit
formal. To justify \eqref{012601-22} rigorously one can 
 modify \eqref{022601-22} as follows:   $\delta(\cdot)$ is replaced by its
approximation, for example
\begin{equation}
\label{delta1}
Nf_*\left(\frac{t-t'}{N}\right)=\frac{1}{2\pi}\int_{\bbR}e^{i\beta(t-t')}\exp\left\{-\frac{\beta^2}{2N}\right\}d\beta,
\end{equation}
when $N\to+\infty$.
Here $f_*(t)=(2\pi)^{-1/2}e^{-t^2/2}$ is the density of the standard
normal distribution. Formula  \eqref{012601-22} is then a consequence
of the passage with $N$ to infinity and an application of the Lebesgue
dominated convergence theorem.
\end{remark}}

Integrating out $s,t$ and $s',t'$ variables we obtain
\begin{align*}
&
\widehat{ w}_{\eps}^{(0)} (\la,\eta,k)=\frac{ \eps T\ga }{(2\pi)\la}\left(1-\frac{1}{2\mu}\right)\int_{\bbR}\Big\{\eps\la/2 -i\om\left(k-\frac{\eps\eta}{2}\right)
  - i\beta
  \Big\}^{-1}\Big\{i\om\left(k+\frac{\eps\eta}{2}\right) +\eps\la/2 +
  i\beta \Big\}^{-1} \\
&
\times \tilde g(\eps\la/2 - i\beta) \tilde g(\eps\la/2 + i\beta) d\beta.
\end{align*}
Change variables
$
\eps\beta':=\beta+\om\left(k-\frac{\eps\eta}{2}\right)
$
and obtain, cf \eqref{nuk},
\begin{align*}
&
\widehat{ w}_{\eps}^{(0)} (\la,\eta,k)=\frac{  T\ga }{(2\pi)\la}\left(1-\frac{1}{2\mu}\right)\int_{\bbR}\Big\{\la/2 
  - i\beta  \Big\}^{-1}\Big\{i \delta_\eps\om(k;\eta)+\la/2 +
  i\beta \Big\}^{-1} \\
&
\times\tilde g\left(\eps\la/2 - i\eps\beta+i
  \om\left(k-\frac{\eps\eta}{2}\right)\right) \tilde g\left(\eps\la/2 +
  i\eps\beta-i \om\left(k+\frac{\eps\eta}{2}\right)\right) d\beta.
\end{align*}
Here
\begin{equation}
\label{deom}
\delta_\eps\om(k;\eta):=\eps^{-1}\Big[\om\left(k+\frac{\eps\eta}{2}\right)-\om\left(k-\frac{\eps\eta}{2}\right)\Big].
\end{equation}
Therefore
\begin{equation}
\label{011112-20}
\lim_{\eps\to0+}\widehat{ w}_{\eps}^{(0)} (\la,\eta,k)=
\frac{  T\ga|\nu(k)|^2 }{(2\pi)\la}\left(1-\frac{1}{2\mu}\right)\int_{\bbR}\Big\{\la/2 
  - i\beta  \Big\}^{-1}\Big\{i \om'(k)\eta+\la/2 +
  i\beta \Big\}^{-1} d\beta.
\end{equation}
To integrate
 out the $\beta$ variable we use the Cauchy  integral formula that in our
 context reads
\begin{equation}
\label{CF}
\frac{1}{2\pi}\int_{\bbR}\frac{
  f(i\beta)d\beta}{z-i\beta}= f(z),\quad z\in \mathbb C_+.
\end{equation}
It is
valid for any holomorphic  function $f$ on the right half-plane $\mathbb C_+$ that belongs to the Hardy class $H^p(\mathbb C_+)$
for some $p\ge 1$,
see e.g. \cite[p. 113]{koosis}. 
Applying the formula we get
\begin{equation}
\label{020112-20}
\lim_{\eps\to0+}\widehat{ w}_{\eps}^{(0)} (\la,\eta,k)
 =
\frac{
  \ga
  T|\nu(k)|^2}{\la(\la+i\om'(k)\eta)}\left(1-\frac{1}{2\mu}\right) .
\end{equation}

\subsection*{Computation of $\widehat{ w}_{\eps}^{(n)} (\la,\eta,k)
  $ for $n\ge1$}

Change variables
$$
\tau_0:=t-s_1,\ldots, \tau_n:=s_n-s_{n+1}(=s_n)
$$
in  \eqref{011709-20}. As a result we get
\begin{align*}
&\widehat{ w}_{\eps}^{(n)}
  (\la,\eta,k)=\frac{ \eps T\ga }{\la}\left(1-\frac{1}{2\mu}\right)\left(\frac{\ga}{\mu}\right)^n
\int_0^{+\infty}
e^{-\eps\la t/2}dt\int_{[0,+\infty)^{n+1}}d\tau_{0,n}\exp\left\{-\eps\la\left(\tau_0+\ldots+\tau_n\right)/2\right\}\\
&
\times\delta\left(t-\tau_0-\ldots-\tau_n\right)
  \phi^\star\left(\tau_0,k-\frac{\eps\eta}{2}\right)
  \phi\left(\tau_0,k+\frac{\eps\eta}{2}\right) \prod_{j=1}^{n}\left(J\star
  g(\tau_{j}) \right)^2 .
\end{align*}
Here $d\tau_{0,n}:=d\tau_0\ldots d\tau_n$. Using \eqref{delta} for each variable $t$ and $\tau_j$, $j=0,\ldots,n$,  we can further write
\begin{align*}
&\widehat{ w}_{\eps}^{(n)}
  (\la,\eta,k)=\frac{ \eps T\ga }{(2\pi)^{n+2}\la}\left(1-\frac{1}{2\mu}\right)\left(\frac{\ga}{\mu}\right)^n
\int_{\bbR}d\beta
  \int_{\bbR^{n+1}}d\beta_{0,n}\int_{[0,+\infty)^{2n+2}}d\tau_{0,n} d\tau'_{0,n} \\
&
\times \int_0^{+\infty}
e^{-(\eps\la /2-i\beta )t}dt
  \prod_{j=0}^n\exp\left\{-(\eps\la/4+i\beta/2+i\beta_j)\tau_j\right\}\prod_{j=0}^n\exp\left\{-(\eps\la/4+i\beta/2-i\beta_j)\tau_j'\right\}\\
&
\times
  \phi^\star\left(\tau_0,k-\frac{\eps\eta}{2}\right)
  \phi\left(\tau_0',k+\frac{\eps\eta}{2}\right) \prod_{j=1}^{n}\left(J\star
  g(\tau_{j}) \right) \prod_{j=1}^{n}\left(J\star
  g(\tau_{j}') \right).
\end{align*}
To abbreviate  we have used the notation $d\beta_{0,n}:=d\beta_0\ldots d\beta_n$ and analogously for the
remaining variables.

Integrating the $t$, $\tau$ variables and its primed counter-parts we get
\begin{align*}
&\widehat{ w}_{\eps}^{(n)}
  (\la,\eta,k)=\frac{ \eps T\ga }{(2\pi)^{n+2}\la}\left(1-\frac{1}{2\mu}\right)\left(\frac{\ga}{\mu}\right)^n
\int_{\bbR}\frac{d\beta}{\eps\la /2-i\beta} \int_{\bbR^{n+1}}d\beta_{0,n}\\
&\times \frac{\tilde g (\eps\la/4+i\beta_0+i\beta/2)}{\eps\la/4+i\Big(\beta_0+\beta/2-\om\left(k-\frac{\eps\eta}{2}\right)\Big)}\times \frac{\tilde g (\eps\la/4-i\beta_0+i\beta/2)}{\eps\la/4+i\Big(\beta/2-\beta_0+\om\left(k+\frac{\eps\eta}{2}\right)\Big)}\\
&
\times
  \prod_{j=1}^{n}\tilde J\tilde
  g (\eps\la/4+i\beta/2+i\beta_j) \prod_{j=1}^{n}\tilde J\tilde
  g (\eps\la/4+i\beta/2-i\beta_j).
\end{align*}
We integrate the $\beta$ variable using the Cauchy integral formula
\eqref{CF} and get
\begin{align*}
&\widehat{ w}_{\eps}^{(n)}
  (\la,\eta,k)=\frac{ \eps T\ga }{(2\pi)^{n+1}\la}\left(1-\frac{1}{2\mu}\right)\left(\frac{\ga}{\mu}\right)^n
\int_{\bbR^{n+1}}d\beta_{0,n}\prod_{j=1}^{n}\tilde J\tilde
  g (\eps\la/2+i\beta_j) \prod_{j=1}^{n}\tilde J\tilde
  g (\eps\la/2-i\beta_j)\\
&\times \frac{\tilde g (\eps\la/2+i\beta_0)}{\eps\la/2+i\Big(\beta_0-\om\left(k-\frac{\eps\eta}{2}\right)\Big)}\times \frac{\tilde g (\eps\la/2-i\beta_0)}{\eps\la/2+i\Big(-\beta_0+\om\left(k+\frac{\eps\eta}{2}\right)\Big)}
 .
\end{align*}
Change of variables
$\eps\beta_0':=\beta_0-\om\left(k-\frac{\eps\eta}{2}\right)$ and obtain
\begin{align*}
&\widehat{ w}_{\eps}^{(n)}
  (\la,\eta,k)=\frac{ T\ga }{(2\pi)^{n+1}\la}\left(1-\frac{1}{2\mu}\right)\left(\frac{\ga}{\mu}\right)^n
\int_{\bbR^{n+1}}d\beta_{0,n}\prod_{j=1}^{n}\tilde J\tilde
  g (\eps\la/2+i\beta_j) \prod_{j=1}^{n}\tilde J\tilde
  g (\eps\la/2-i\beta_j)\\
&\times \frac{\tilde g (\eps\la/2+i\eps\beta_0+i
  \om\left(k-\frac{\eps\eta}{2}\right))}{\la/2+i\beta_0}\times
  \frac{\tilde g (\eps\la/2-i\eps\beta_0-i
  \om\left(k-\frac{\eps\eta}{2}\right))}{\la/2+i\left(-\beta_0+\delta_\eps\om(k;\eta)\right)}.
\end{align*}
\blue{According to Lemma \ref{lm011811-20} we have $\tilde J\tilde
  g\in H^2(\mathbb C_+)$, therefore, see e.g. \cite[Theorem 19.2]{rudin},
 $$
\frac{\ga}{2\pi}\int_{\bbR}|\tilde J\tilde
  g ( \la+i\beta) |^2d\beta\le \frac{\ga}{2\pi}\int_{\bbR}|\tilde J\tilde
  g ( i\beta) |^2d\beta=\Gamma<\frac{1}{2}\le \mu,\quad {\rm Re}\,\la>0.
$$
The last estimate follows from \eqref{Gamma-a}. In particular, there
exists a constant $C>0$ such that 
\begin{equation}
\label{032501-22}
|\widehat{ w}^{(n)}_\eps
  (\la,\eta,k)|\le C\left(\frac{\Gamma}{\mu}\right)^n,\quad n\ge0,\,\eps>0,\, \la\in\mathbb C_+\,\mbox{ and }(\eta,k)\in\bbR\times\bbT.
\end{equation}
This proves that the validity of \eqref{021601-19z}   for all $\la\in\mathbb C_+$.}

Furthermore,
\begin{align*}
&\widehat{ w}^{(n)}
  (\la,\eta,k):=\lim_{\eps\to0+}\widehat{ w}_{\eps}^{(n)}
  (\la,\eta,k)\\
&
=\frac{  \ga T}{\la(2\pi)}\left(\frac{\Gamma}{\mu}\right)^n|\nu(k)|^2 \left(1-\frac{1}{2\mu}\right)
\int_{\bbR}\frac{d\beta_{0}}{(\la/2+i\beta_0)
  \big\{\la/2+i\big[-\beta_0+\om'(k)\eta\big]\big\}}.
\end{align*}
Here $\Gamma$ is given by \eqref{Gamma}.
Integrating the $\beta_0$ variable out, using again \eqref{CF}, we get
$$
\widehat{ w}^{(n)}
  (\la,\eta,k)
=
\frac{  \ga T |\nu(k)|^2}{\la(\la+i\om'(k)\eta)}\left(\frac{\Gamma}{\mu}\right)^n \left(1-\frac{1}{2\mu}\right).
$$



Using \eqref{032501-22}, by the dominated convergence theorem, we conclude that
\begin{equation}
\label{040112-20}
\widehat{ w} (\la,\eta,k)=\sum_{n=0}^{+\infty} \widehat{
  w}^{(n)} (\la,\eta,k) 
\end{equation}
and formula \eqref{021601-19zz} follows.\qed

\bigskip


\section{The case $T=0$ and non-zero initial data}

\label{sec5-1}

Here, as in Section \ref{sec5}, we assume that $T=0$ and the initial
data need not be null, and satisfies the assumptions made in Sections
\ref{sec2.4.3} and \ref{sec2.4.4}. The
solution $\hat\psi(t,k)$  is then described by the expansion
\eqref{momentum-3a} and \eqref{eq:sol-xx}. 
Using the same argument as in the proof of Lemma \ref{lm020112-20} we
obtain the following Poisson chaos expansion for the momentum at $x=0$
and the Fourier transform of the wave function
\begin{equation}
 \label{momentum-3f}
 {\frak p}_0(t) 
=g\star{\frak p}_0^0(t)+\sum_{n=1}^{+\infty}\left(-\frac{1}{\mu}\right)^{n}\int_{\Delta_n(t)}\prod_{j=1}^nJ\star
 g(s_{j-1}-s_j)g\star {\frak p}_0^0(s_n) d\tilde N( \ga\mu s_1)\ldots
 d\tilde N( \ga\mu s_n),
\end{equation}
and
\begin{equation}
  \label{eq:sol1x}
  \begin{split}
  &  \hat\psi(t,k) 
=e^{-i\omega(k) t} \hat\psi(0,k)-i\ga\int_0^t \phi(t-s,k){\frak
  p}_0^0(s)ds \\
&
+ i\sum_{n=1}^{+\infty}\left(-\frac{1}{\mu}\right)^{n}\int_{\Delta_n(t)}\phi(t-s_1,k)
 \prod_{j=1}^{n-1}J\star
 g(s_{j}-s_{j+1})   g\star{\frak p}_0^0(s_n) d\tilde N( \ga \mu s_1)\ldots
 d\tilde N( \ga \mu s_n),
  \end{split}
\end{equation}
where ${\frak p}_0^0(\cdot)$ is given by \eqref{eq:bessel0}. {In
  light of \eqref{Gamma-a}  both of
these expansions are valid for any $\mu\ge1/2$.}

On the other hand from \eqref{basic:sde:2aas}, with $\tilde
  p(t)\equiv0$, we obtain the following equation on the Fourier-Wigner
  function $W_\eps(t,\eta,k)$
 \begin{eqnarray}
\label{exp-wigner-eqt0}
&&\partial_t\widehat
   W_\eps(t,\eta,k)+i\delta_{\eps}\om(k;\eta)\widehat
   W_\eps(t,\eta,k) =\frac{\ga}{2\mu}\bbE\left[{\frak p}_0^2\left(\frac{t}{\eps}\right) \right]\\
&&
+
\frac{i\ga}{2}\left\{\bbE\left[\hat\psi\left(\frac{t}{\eps},k+\frac{\eps \eta}{2}\right) {\frak p}_0\left(\frac{t}{\eps}\right)\right]
-\bbE\left[\hat\psi^*\left(\frac{t}{\eps},k-\frac{\eps \eta}{2}\right) {\frak p}_0\left(\frac{t}{\eps}\right)\right]\right\}\notag
\end{eqnarray}
Taking the Laplace transform on both sides we arrive at
\begin{equation}
\label{exp-wigner-eqt}
\begin{split}
&\left(\la+i\delta_{\eps}\om(k;\eta) \right)\widehat
   w_\eps(\la,\eta,k)
 =W_\eps(0,\eta,k)\\
&
\qquad \qquad \qquad \qquad +\frac{\ga}{\mu}{\frak e}_\eps(\la)-\frac{\ga}{2}\left[{\frak
    d}_\eps\left(\la,k-\frac{\eps \eta}{2}\right)+{\frak
    d}_\eps^\star\left(\la,k+\frac{\eps \eta}{2}\right)\right],
\end{split}
\end{equation}
where
\begin{align}
\label{020712-20}
&
{\frak e}_\eps(\la):=\frac{\eps}{2}\int_0^{+\infty}e^{-\la\eps
   t}\bbE\left[{\frak p}_0^2\left(t\right) \right]dt\quad\mbox{and}\notag\\
&
\\
&
{\frak d}_\eps(\la,k):=i\eps\int_0^{+\infty}e^{-\la\eps
   t}\bbE\left[\hat\psi^*\left(t,k\right) {\frak p}_0\left(t\right)\right]dt.\notag
\end{align}
In the present section we show the following.
\begin{prop}
\label{prop010412-20}
For any $G\in {\cal S}(\bbR\times\bbT)$ and ${\rm Re}\,\la>0$ we have
$$
\int_{\bbR}\int_{\bbT}\widehat{ w}(\la,\eta,k)G^*(\eta,k)d\eta dk=\lim_{\eps\to0+}\int_{\bbR}\int_{\bbT}\widehat{
  w}_{\eps}(\la,\eta,k) G^*(\eta,k)d\eta dk,
$$
where
\begin{equation}
\label{010412-20}
\begin{split}
&\widehat{ w}(\la,\eta,k)=\frac{\widehat
  W(0,\eta,k)}{\la+i\om'(k)\eta}+\frac{\ga |\nu(k)|^2}{2(1-\Gamma/\mu)( \la+i\om'(k)\eta)}\int_{\bbR}\int_{\bbT}\frac{\widehat
  W(0,\eta,\ell)|\nu(\ell)|^2}{\la+i\om'(\ell)\eta} d\eta d\ell\\
&
-\frac{\ga {\rm Re} [\nu(k)]}{\la+i\om'(k)\eta} \int_{\bbR\times\bbT}
 \frac{\widehat W(0,\eta',k)}{\la+i\om'(k)\eta'} 
 d\eta'
\\
&
+\frac{\gamma {\frak g}(k)}{4(\la+i\om'(k)\eta)}\int_{\bbR\times\bbT}\frac{\widehat
  W(0,\eta',k)d\eta' }{\la+i\om'(k)\eta'} 
+\frac{\gamma {\frak g}(k)}{4(\la+i\om'(k)\eta)}\int_{\bbR\times \bbT}\frac{\widehat
  W(0,\eta',-k)d\eta' }{\la-i\om'(k)\eta'} .
\end{split}
\end{equation}
\end{prop}
The proof of the proposition is carried out throughout
Sections \ref{sec5.2} - \ref{sec5.3a}.

\subsection{Asymptotics of ${\frak e}_\eps(\la)$}

\label{sec5.2}

\begin{prop}
\label{prop010212-20}
Under the assumption about the initial data made in Sections
\ref{sec2.4.3} and \ref{sec2.4.4}  we have
\begin{equation}
\label{E}
\lim_{\eps\to0+}{\frak e}_\eps(\la)=\frac{1}{2(1-\Gamma/\mu)}
  \int_{\bbR}\int_{\bbT}\frac{\widehat W(0,\eta,\ell)|\nu(\ell)|^2}{\la+i\om'(\ell)\eta} d\eta d\ell.
\end{equation}
\end{prop}
\proof
From \eqref{momentum-3f} we get
\begin{equation}
\begin{split}
 \label{momentum-3fe}
\bbE\left[{\frak p}_0^2\left(t\right) \right]
=\bbE[g\star{\frak p}_0^0(t)]^2+\sum_{n=1}^{+\infty}\left(\frac{\ga}{\mu}\right)^n\int_{\Delta_n(t)}\prod_{j=1}^n(J\star
 g)^2(s_{j-1}-s_j)\bbE[g\star {\frak p}_0^0(s_n)]^2 ds_1\ldots
 ds_n.
\end{split}
 \end{equation}
\blue{Arguing as in the proof of Proposition \ref{prop010112-20}  we
conclude that for $\la\in\mathbb C_+$}
\begin{equation}
\begin{split}
 \label{momentum-3fen}
&{\frak e}_\eps(\la)=\sum_{n=0}^{+\infty}E_n^{(\eps)}(\la),\quad
\mbox{where}
\\
&
E_0^{(\eps)}(\la):=\frac{\eps}{2}\int_0^{+\infty}e^{-\la\eps
   t}\bbE[g\star{\frak p}_0^0(t)]^2dt,\\
&
E_n^{(\eps)}(\la):= \frac{\eps }{2}\left(\frac{\ga}{\mu}\right)^n\int_0^{+\infty}e^{-\la\eps
   t}dt\int_{\Delta_n(t)}(J\star
 g)^2(t-s_1) \prod_{j=1}^{n-1}(J\star
 g)^2(s_{j}-s_{j+1})\\
 &\qquad\qquad\qquad\qquad\qquad\qquad\qquad\qquad\times\bbE[g\star {\frak p}_0^0(s_n)]^2 ds_1\ldots
 ds_n.
\end{split}
 \end{equation}

\subsubsection*{Asymtotics of $E_0^{(\eps)}(\la)$}

\label{sec5.2.1}
Using \eqref{delta} we can write
\begin{align*}
&
E_0^{(\eps)}(\la)
=-\frac{\eps}{2^4\pi}\int_{\bbT^2}dk dk'\int_0^{+\infty}dt\int_0^{+\infty}dt'e^{-\la\eps
   (t+t')/2}\int_0^t \int_0^{t'}g(d\si) g(d\si') \int_{\bbR}d\beta e^{i\beta(t-t')}\\
&
\times \bbE\left\{\left\{e^{-i\om(k)(t-\si)}\hat\psi(k)- e^{i\om(k)(t-\si)}\hat\psi^*(k)\right\}\left\{e^{-i\om(k')(t'-\si')}\hat\psi(k')- e^{i\om(k')(t'-\si')}\hat\psi^*(k')\right\}\right\}.
\end{align*}
Thanks to \eqref{null} we can write
\begin{align*}
&
E_0^{(\eps)}(\la)
=\frac{\eps}{2^3\pi}\int_{\bbT^2}dk dk'\int_0^{+\infty}dt\int_0^{+\infty}dt'e^{-\la\eps
   (t+t')/2}\int_0^t \int_0^{t'}g(d\si) g(d\si') \int_{\bbR}d\beta e^{i\beta(t-t')}\\
&
\times \exp\left\{i\om(k')(t'-\si')-i\om(k)(t-\si)\right\}\bbE\left\{\hat\psi(k) \hat\psi^*(k')\right\}.
\end{align*}
Integrating out the $t$ and $t'$ variables we get
\begin{align*}
E_0^{(\eps)}(\la)
=\frac{\eps}{2^3\pi}\int_{\bbT^2}\int_{\bbR}
  \frac{|\tilde g(\la\eps/2-i\beta)|^2 }{\la\eps/2-i\beta+i\om(k)}\cdot\frac{\bbE\left\{\hat\psi(k) \hat\psi^*(k')\right\}}{\la\eps/2+i\beta-i\om(k')}dkdk' d\beta 
.
\end{align*}
Next we change variables $\eps\beta':=\beta-\om(k')$, which leads to
\begin{align}
\label{E0}
E_0^{(\eps)}(\la)
=\frac{1}{2^3\pi}\int_{\bbT^2}\int_{\bbR} 
  \frac{|\tilde g(\la\eps/2-i\eps\beta+i\om(k))|^2}{\la/2-i\beta+i\eps^{-1}[\om(k)-\om(k')]}\cdot\frac{\bbE\left\{\hat\psi(k) \hat\psi^*(k')\right\}}{\la/2+i\beta}dk dk' d\beta
. 
\end{align}
Change variables $(k,k')\mapsto (\eta,\ell)$, by letting
\begin{equation}
\label{k-eta}
k:=\ell+\frac{\eps \eta}{2},\qquad k':=\ell-\frac{\eps \eta}{2}.
\end{equation}
The image of $\bbT^2$ under this mapping is 
\begin{equation}
\label{T2e}
T^2_\eps:=\left[(\eta,\ell):\,|\eta|\le
\frac{1}{\eps},\,|\ell|\le\frac{1-\eps |\eta|}{2}\right]
\subset\bbT_{2/\eps}\times \bbT.
\end{equation}
Then, cf \eqref{deom},
$$
E_0^{(\eps)}(\la)
=\frac{1}{2^2\pi}\int_{T^2_\eps}\widehat W_\eps(0,\eta,\ell) d\eta d\ell \int_{\bbR}d\beta
  \frac{ \left|\tilde g\left(\la\eps/2-i\eps\beta+i\om(\ell+\frac{\eps \eta}{2})\right)\right|^2 }{\left(\la/2-i\beta+i\delta_\eps\om(\ell,\eta)\right)\left(\la/2+i\beta\right)}.
 $$
Using estimates  \eqref{011812aa}, \eqref{011812c} and the Cauchy
formula \eqref{CF},  we obtain
\begin{align}
\label{010212-20}
&
\lim_{\eps\to0+}E_0^{(\eps)}(\la)=\frac{1}{2^2\pi}\int_{\bbR}d\eta \int_{\bbT}d\ell \int_{\bbR}d\beta 
  \frac{\left|\nu(\ell)\right|^2 \widehat
  W(0,\eta,\ell)}{\left(\la/2-i\beta+i\om'(\ell)\eta\right)\left(\la/2+i\beta\right)}
\notag\\
&
=\frac12 \int_{\bbR} \int_{\bbT}
  \frac{\widehat
  W(0,\eta,\ell) \left|\nu(\ell)\right|^2 }{\la+i\om'(\ell)\eta}d\eta d\ell.
\end{align}

\subsubsection*{Asymptotics  of $E_n^{(\eps)}(\la)$ for $n\ge1$}

Using \eqref{eq:bessel0} and \eqref{null} we get
\begin{align*}
&
E_n^{(\eps)}(\la)
=\frac{\eps }{2^2}\left(\frac{\ga}{\mu}\right)^n\int_{\bbT}dk \int_{\bbT}dk' \bbE\left\{\hat\psi(k) \hat\psi^*(k')\right\}\int_0^{+\infty}e^{-\la\eps
   t}dt\int_{\Delta_n(t)} (J\star
 g)^2(t-s_1) dt
\\
&
\times \prod_{j=1}^{n-1}(J\star
 g)^2(s_{j}-s_{j+1})\int_0^{s_n} \int_0^{s_n}g(d\si_1) g(d\si'_1) \exp\left\{i\om(k')(s_n-\si'_1)-i\om(k)(s_n-\si_1)\right\}.
\end{align*}
We substitute $\tau_j:=s_{j}-s_{j+1}$, $j=0,\ldots,n$, with $s_0:=t$
  and  $s_{n+1}:=0$, and then use \eqref{delta} to double 
  variables $\tau_j$ and $\tau_j'$. In this way we obtain
\begin{align*}
&
E_n^{(\eps)}(\la)
=\frac{\eps
  }{2^2(2\pi)^{n+3}}\left(\frac{\ga}{\mu}\right)^n\int_{\bbR^2}d\beta
  d\beta' \int_{\bbT^2}dk dk'\int_{(0,+\infty)^{2}}dt dt'e^{-\la\eps
   (t+t')/4}\bbE\left\{\hat\psi(k) \hat\psi^*(k')\right\}\\
&
\times 
\int_{(0,+\infty)^{n+1}}d\tau_{0,n}\int_{(0,+\infty)^{n+1}}d\tau_{0,n}'
\int_{\bbR^{n+1}}d\beta_{0,n} \prod_{j=0}^{n}
  e^{i\beta_j(\tau_j-\tau_j')}\\
&
\times  \exp\left\{-\la\eps
   \left(\sum_{j=0}^n\tau_j\right)/4\right\}\exp\left\{-\la\eps
   \left(\sum_{j=0}^n\tau_j'\right)/4\right\}\exp\left\{i\beta\left(t-\sum_{j=0}^n\tau_j\right)\right\}\exp\left\{i\beta'\left(t'-\sum_{j=0}^n\tau_j'\right)\right\} 
\\
&
\times\prod_{j=0}^{n-1}(J\star
 g)(\tau_j) \prod_{j=0}^{n-1}(J\star
 g)(\tau_j')
\int_0^{\tau_n} \int_0^{\tau_n'}g(d\si) g(d\si') \exp\left\{i\om(k')(\tau_n'-\si')-i\om(k)(\tau_n-\si)\right\}.
\end{align*}
To abbreviate we have used the notation
$d\tau_{0,n}=d\tau_0\ldots d\tau_n$,
$d\beta_{0,n}=d\beta_0\ldots d\beta_n$ and similarly for the prime variables.
Integrating out the $t$, $\tau$ variables and their prime counterparts we get
\begin{align}
\label{En}
&
E_n^{(\eps)}(\la)
=\frac{\eps}{2^2(2\pi)^{n+3}}\left(\frac{\ga}{\mu}\right)^n\int_{\bbR^2}d\beta
  d\beta' \int_{\bbT^2}dk dk'
\int_{\bbR^{n+1}}d\beta_{0,n}   \bbE\left\{\hat\psi(k) \hat\psi^*(k')\right\}
 \notag\\
&
\times\prod_{j=0}^{n-1}(\tilde J\tilde
 g)\left(\la\eps/4-i\beta_j+i\beta\right) \prod_{j=0}^{n-1}(\tilde J\tilde
 g)\left(\la\eps/4+i\beta_j+i\beta'\right)
\notag\\
&
\frac{1}{\la\eps/4-i\beta}\cdot\frac{1}{\la\eps/4-i\beta'}\cdot\frac{\tilde g(\la\eps/4-i\beta_n+i\beta)}{\la\eps/4-i\beta_n+i\beta+i\om(k)}\cdot \frac{\tilde g(\la\eps/4+i\beta_n+i\beta')}{\la\eps/4+i\beta_n+i\beta'-i\om(k')}.
\end{align}
Change variables $k,k'$ according to \eqref{k-eta} and 
\begin{align*}
\eps\beta_n:=\beta_n-\om(k'),\quad \eps\beta:=\beta,\quad \eps \beta':=\beta'
\end{align*}
we obtain
\begin{align*}
&
E_n^{(\eps)}(\la)
=\frac{1}{2(2\pi)^{n+3}}\left(\frac{\ga}{\mu}\right)^n
  \int_{\bbR^2}\frac{d\beta  d\beta'}{(\la/4-i\beta)(\la/4-i\beta')} \int_{T^2_\eps}\widehat W_\eps(0,\eta,\ell)
d\eta d\ell
 \\
&
\times \int_{\bbR^{n+1}}d\beta_{0,n}\prod_{j=0}^{n-1}(\tilde J\tilde
 g)\left(\la\eps/4-i\beta_j+i\eps\beta\right) \prod_{j=0}^{n-1}(\tilde J\tilde
 g)\left(\la\eps/4+i\beta_j-i\eps\beta'\right)
\\
&
\times\frac{\tilde g(\la\eps/4-i\eps\beta_n-i\om(\ell-\eps\eta/2)+i\eps\beta) }{\la/4-i\beta_n+i\beta+i\delta_\eps\om(\ell,\eta)}\cdot \frac{\tilde g(\la\eps/4+i\eps\beta_n+i\om(\ell-\eps\eta/2)+i\eps\beta')}{\la/4+i\beta_n+i\beta'}.
\end{align*}
Hence
\begin{align}
\label{020212-20}
\lim_{\eps\to0+}E_n^{(\eps)}(\la)=\frac{\Gamma^n}{2\mu^n}
  \int_{\bbR}\int_{\bbT}\frac{\widehat W(0,\eta,\ell)|\nu(\ell)|^2}{\la+i\om'(\ell)\eta} d\eta d\ell.
\end{align}
The conclusion of the proposition then follows from an application of
the   dominated convergence theorem to the series appearing in \eqref{momentum-3fen}, as $\Gamma/\mu\in(0,1)$.\qed

\subsection{Asymptotics of the term involving ${\frak d}_\eps(\la)$}

\label{sec5.3}

Invoking \eqref{exp-wigner-eqt} we wish to calculate the limit
$\lim_{\eps\to0+}{\frak L}_\eps$, where
\begin{equation}
\label{070212-20}
{\frak L}_\eps:=\int_{\bbT}\int_{\bbR}\left[{\frak
    d}_\eps\left(\la,k-\frac{\eps \eta}{2}\right)+{\frak
    d}_\eps^\star\left(\la,k+\frac{\eps
      \eta}{2}\right)\right]\frac{G^\star(\eta,k)d\eta dk}{\la+i\delta_{\eps}\om(k;\eta) },
\end{equation}
for any $G\in {\cal S}(\bbR\times\bbT)$.

Taking into account \eqref{momentum-3f} and \eqref{eq:sol1x} 
we get
\begin{equation}
\label{D-eps}
{\frak d}_\eps(\la,k)=\sum_{n=0}^{+\infty}D_n^{\eps}(\la,k),
\end{equation}
where
\begin{equation}
\label{D-eps-0}
D_0^{\eps}(\la,k)=D_{0,1}^{\eps}(\la,k)+D_{0,2}^{\eps}(\la,k)
\end{equation}
and
\begin{align}
\label{030212-20}
&
D_{0,1}^{\eps}(\la,k):=i\eps\int_0^{+\infty}e^{-\la\eps
   t}e^{i\omega(k) t}\bbE\left[ \hat\psi^\star(0,k) g\star{\frak p}_0^0(t)\right]dt,\notag\\
&
D_{0,2}^{\eps}(\la,k):=-\eps\ga\int_0^{+\infty}e^{-\la\eps
   t}dt\int_0^t \phi^\star(t-s,k)\bbE\left[ {\frak
  p}_0^0(s) g\star{\frak p}_0^0(t)\right]ds,\notag\\
&
D_n^{\eps}(\la,k):=\eps \left(\frac{\ga}{\mu}\right)^n \int_0^{+\infty}e^{-\la t}dt\int_{\Delta_n(t)}\phi^*(t-s_1,k) (J\star
 g)(t-s_1)
 \\
&
\times\prod_{j=1}^{n-1}(J\star
 g)^2(s_{j}-s_{j+1})  \bbE\left[ \left(g\star{\frak p}_0^0(s_n) \right)^2\right]ds_1\ldots
 ds_n,\,n\ge1\notag.
\end{align}
Accordingly we can write ${\frak L}_\eps=\sum_{n=0}^{+\infty}{\frak
  L}_\eps^{(n)}$, where
\begin{equation}
\label{070212-20n}
{\frak
  L}_\eps^{(n)}:=\int_{\bbR}\int_{\bbT}\left[D_n^{\eps}\left(\la,k-\frac{\eps
      \eta}{2}\right)+(D_n^{\eps})^*\left(\la,k+\frac{\eps
      \eta}{2}\right)\right]\frac{G^\star(\eta,k)}{\la+i\delta_{\eps}\om(k;\eta)
}d\eta dk.
\end{equation}

\subsubsection{Computation of $D_{0,1}^{\eps}(\la,k)$}
The term $D_{0,1}^{\eps}(\la,k)$ coincides with ${\frak
  d}_\eps^1(\la,k)$ defined in \cite[formulas (5.6) and  (5.7)]{kors}.
Therefore, see \cite[Lemma 5.1]{kors}, we have the following result.
\begin{lm} \label{lem-feb1504}
For any test function $G\in {\cal S}(\bbR\times\bbT)$ and
  $\la>0$ we have
\begin{equation}
\label{020811}
\begin{split}&
-\frac{\ga}{2}
\lim_{\eps\to0+}\int_{\bbR\times\bbT}\frac{G^*(\eta,k)}{\la+i\delta_\eps\om(k,\eta)}\left\{D_{0,1}^{\eps}\left(\la,k-\frac{\eps\eta}{2}\right)+\left(D_{0,1}^{\eps}\right)^*\left(\la,k+\frac{\eps\eta}{2}\right)\right\}d\eta
  dk\\
&
=
 -\ga \int_{\bbR\times\bbT}{\rm Re} [\nu(k)]
 \frac{\widehat W(0,\eta',k)}{\la+i\om'(k)\eta'} 
 \left\{\int_{\bbR}\frac{G^*(\eta,k)}{\la+i\om'(k)\eta}d\eta\right\}dk
 d\eta'.
\end{split}
\end{equation}
\end{lm}

\subsubsection{Asymptotics of $D_{0,2}^{\eps}(\la,k)$}

Using \eqref{phi-t1} we can write
\begin{align*}
D_{0,2}^{\eps}(\la,k)
=-\eps\ga\int_0^{+\infty}e^{-\la\eps
   t}dt\int_0^tds\exp\left\{i\om(k)(t-s)\right\}
 \bbE\left[ g\star{\frak
  p}_0^0(s) g\star{\frak p}_0^0(t)\right]
\end{align*}
The expression for $D_{0,2}^{\eps}(\la,k)$ is therefore identical with ${\frak
    d}_\eps^{2}\left(\la,k\right)$ defined by \cite[formulas (5.6) and  (5.7)]{kors}. We have therefore,
  see \cite[Lemma 5.2]{kors}.
\begin{lemma}\label{lem-feb1502}
For any $\la>0$ and $G\in {\cal S}(\bbR\times \bbT)$ we have
\begin{eqnarray}
\label{021511}
&&-\frac{\ga}{2}\lim_{\eps\to 0}\int_{\bbR\times\bbT}\left[ D_{0,2}^{\eps}\left(\la,k-\frac{\eps \eta}{2}\right)+\left(D_{0,2}^{\eps}\right)^*\left(\la,k+\frac{\eps
   \eta}{2}\right)\right]\frac{\hat G^*(\eta,k) d\eta dk}{\la+ i\delta_\eps\om(k,
\eta)} \notag\\
&&
=
\frac{\gamma}{4}\int_{\bbR\times\bbT}\frac{\fgeeszett(k)\widehat
  W(0,\eta',k)d\eta' dk}{\la+i\om'(k)\eta'} \int_{\bbR}
  \frac{\hat G^*(\eta,k)d\eta}{\la+i\om'(k)\eta}
\\
&&
+\frac{\gamma}{4}\int_{\bbR\times \bbT}\frac{\fgeeszett(k)\widehat
  W(0,\eta',-k)d\eta' dk}{\la-i\om'(k)\eta'} 
  \int_{\bbR}\frac{\hat G^*(\eta,k)d\eta}{\la+i\om'(k)\eta}.\nonumber
\end{eqnarray}
\end{lemma}
Summarizing, taking into account definitions \eqref{033110}, we have
\begin{equation}
\label{080212-20}
-\frac{\ga}{2}\lim_{\eps\to 0}{\frak L}_\eps^{(0)}=\frac{\big(p_+(k)-1\big)|\bar\om'(k)|}{\la+i\om'(k)\eta}\int_{\bbR\times\bbT}\frac{\widehat
  W(0,\eta',k)d\eta' }{\la+i\om'(k)\eta'} 
+\frac{p_-(k)|\bar\om'(k)|}{\la+i\om'(k)\eta}\int_{\bbR}\frac{\widehat
  W(0,\eta',-k)d\eta' }{\la-i\om'(k)\eta'} .
\end{equation}

\subsubsection{Asymptotics of $\sum_{n=1}^{+\infty}
  D_{n}^{\eps}(\la,k)$}

We prove the following.
\begin{lemma}\label{lm010312-20}
For any $\la>0$   we have
\begin{eqnarray}
\label{050312-20}
&&-\frac{\ga}{2}\lim_{\eps\to 0}\sum_{n=1}^{+\infty}\int_{\bbR\times\bbT}\left[ D_{n}^{\eps}\left(\la,k-\frac{\eps \eta}{2}\right)+\left(D_{n}^{\eps}\right)^*\left(\la,k+\frac{\eps
   \eta}{2}\right)\right]\frac{\hat G^*(\eta,k) d\eta dk}{\la+ i\delta_\eps\om(k,
\eta)} \\
&&
=
-\frac{\ga}{2\mu (1-\Gamma/\mu)} \int_{\bbR\times\bbT}\frac{G^*(\eta,k) [1-|\nu(k)|^2]d\eta dk}{\la+i\om'(k)\eta}
  \int_{\bbR\times\bbT} \frac{|\nu(\ell)|^2\widehat
  W(0,\eta',\ell) d\eta' d\ell}{\la+i\om'(\ell)\eta'}.\nonumber
\end{eqnarray}
\end{lemma}
The proof of the lemma is presented in Section \ref{sec5.2.5}. It requires some auxiliary calculations that
are done in Section \ref{sec5.2.4}.

\subsubsection{Auxiliary calculations}

\label{sec5.2.4}

We suppose that $n\ge1$.
Using the change of variables
$\tau_j:=s_j-s_{j+1}$, $j=0,\ldots,n$, with $s_0:=t$ and $s_{n+1}:=0$
in the last formula of  \eqref{030212-20}
and then \eqref{delta} we get
\begin{align}
\label{030812-20}
& D_n^{\eps}(\la,k)=\frac{\eps }{2\pi} \left(\frac{\ga}{\mu}\right)^n\int_0^{+\infty}e^{-\la \eps t/2}dt
  \int_{\bbR}d\beta\int_{(0,+\infty)^{n+1}}d\tau_{0,n}\exp\left\{i\beta \left(t-\sum_{j=0}^n\tau_j\right)\right\}\\
&
\times \exp\left\{-\la \eps\left(\sum_{j=0}^n\tau_j\right)/2\right\}\phi^*(\tau_0,k) (J\star
 g)(\tau_0)
 \prod_{j=1}^{n-1}(J\star
 g)^2(\tau_{j})  \bbE\left[ \left(g\star{\frak p}_0^0(\tau_n) \right)^2\right],\,n\ge1.\notag
\end{align}
Doubling the   $\tau_j$ variables, via \eqref{delta}, we get
\begin{align*}
& D_n^{\eps}(\la,k)=\frac{\eps }{(2\pi)^{n+2}}\left(\frac{\ga}{\mu}\right)^n \int_0^{+\infty}dt
  \int_{\bbR^{n+2}}d\beta_{0,n}  d\beta \int_{(0,+\infty)^{n+1}}d\tau_{0,n}\int_{(0,+\infty)^{n+1}}d\tau_{0,n}'
\\
&
\times e^{-\la\eps
  t /2}\prod_{j=0}^n \exp\left\{i\beta_j(\tau_j-\tau_j')\right\}
\exp\left\{i\beta
  \left(t-\frac12\sum_{j=0}^n\tau_j-\frac12\sum_{j=0}^n\tau_j'\right)\right\}\\
&
\times \exp\left\{-\la\eps\left(\sum_{j=0}^n\tau_j\right)
  /4\right\}\exp\left\{-\la\eps\left(\sum_{j=0}^n\tau_j'\right)/4\right\}
\\
&
\times\phi^*(\tau_0',k) (J\star
 g)(\tau_0)
 \prod_{j=1}^{n-1}(J\star
 g)(\tau_{j}) \prod_{j=1}^{n-1}(J\star
 g)(\tau_{j}')  \bbE\left[ \left(g\star{\frak p}_0^0(\tau_n) \right)  \left(g\star{\frak p}_0^0(\tau_n') \right)\right].
\end{align*}
Integrating out the $t$, $\tau$ and $\tau'$ variables we get
\begin{align}
\label{050212-20}
&
D_n^{\eps}(\la,k)
=\frac{\eps }{(2\pi)^{n+2}}\left(\frac{\ga}{\mu}\right)^n \int_{\bbR}\frac{d\beta}{\la\eps/2 -i\beta}
\int_{\bbR^{n+1}}d\beta_{0,n} (\tilde J\tilde
 g)\left(\la\eps /4-i\beta_0+i\beta/2 \right) \tilde \phi^*(\la\eps /4-i\beta_0-i\beta/2,k)
\notag \\
&
\times\prod_{j=1}^{n-1}(\tilde J\tilde
 g)\left(\la\eps /4-i\beta_j+i\beta/2\right) \prod_{j=1}^{n-1}(\tilde J\tilde
 g)\left(\la\eps  /4+i\beta_j+i\beta/2\right)
\tilde g(\la\eps /4-i\beta_n+i\beta/2) \tilde
  g(\la\eps/4+i\beta_n+i\beta/2)\\
&
\times \bbE\left[ \tilde {\frak p}_0^0(\la\eps
  /4-i\beta_n+i\beta/2 ) \tilde {\frak p}_0^0(\la\eps
  /4+i\beta_n+i\beta/2 ) \right]. \notag
\end{align}
Here 
$$
\tilde \phi(\la,k)=\frac{\tilde g(\la)}{\la+i\om(k)}
$$
and
 \begin{align*}
\tilde {\frak p}_0^0(\la)
=\frac{1}{2i}\int_{\bbT}\left\{\frac{\hat\psi(\ell)}{\la+i\om(\ell)}-\frac{\hat\psi^*(\ell)}{\la-i\om(\ell)}\right\}d\ell
\end{align*}
are the Laplace transforms of $\phi(t,k)$ and ${\frak p}_0^0(t)$, respectively.

Thanks to \eqref{null} we have
\begin{align*}
&
\bbE\left[\tilde {\frak p}_0^0(\la_1) \tilde {\frak p}_0^0(\la_2)\right]
=\frac{1}{2^2}\int_{\bbT}d\ell \int_{\bbT}d\ell'\left\{\frac{\bbE\left[\hat\psi(\ell) \hat\psi^*(\ell')\right]}{\big(\la_1+i\om(\ell)\big)\big(\la_2-i\om(\ell')\big)}+\frac{\bbE\left[\hat\psi(\ell') \hat\psi^*(\ell)\right]}{\big(\la_1-i\om(\ell)\big)\big(\la_2+i\om(\ell')\big)}\right\}
\end{align*}
Substituting into \eqref{050212-20} we get
\begin{align*}
&
D_n^{\eps}(\la,k)
=\frac{\eps }{2^2(2\pi)^{n+2}}\left(\frac{\ga}{\mu}\right)^n\int_{\bbT^2}d\ell
   d\ell' \int_{\bbR}\frac{d\beta}{\la\eps /2-i\beta} 
\int_{\bbR^{n+1}}d\beta_{0, n} (\tilde J\tilde
 g)\left(\la\eps /4-i\beta_0+i\beta /2\right) \\
&
\times\frac{\tilde g(\la\eps
  /4+i\beta_0+i\beta /2)}{\la\eps /4+i\beta_0+i\beta /2-i\om(k)} \tilde g(\la\eps /4-i\beta_n+i\beta /2) \tilde
  g(\la\eps /4+i\beta_n+i\beta /2)
 \\
&
\times\prod_{j=1}^{n-1}(\tilde J\tilde
 g)\left(\la\eps /4-i\beta_j+i\beta /2\right) \prod_{j=1}^{n-1}(\tilde J\tilde
 g)\left(\la\eps /4+i\beta_j+i\beta /2\right)
\\
&
\times \left\{\frac{\bbE\left[\hat\psi(\ell)
  \hat\psi^*(\ell')\right]}{[\la\eps /4-i\beta_n+i\beta
  /2+i\om(\ell)][\la\eps /4+i\beta_n+i\beta /2-i\om(\ell')]}\right.\\
&
\left.+\frac{\bbE\left[\hat\psi^*(\ell)
  \hat\psi(\ell')\right]}{[\la\eps /4-i\beta_n+i\beta
  /2-i\om(\ell)][\la\eps /4+i\beta_n+i\beta /2+i\om(\ell')]}\right\}.
\end{align*}

Change variables 
$\beta_j':=\beta_j+\beta /2$, $j=0,\ldots,n$ and integrate out the
$\beta$ variable, using \eqref{CF}. We can
write then
\begin{align}
\label{090212-20}
&
D_n^{\eps}(\la,k)
= \frac{1}{4 \mu^n}\left(\frac{\ga}{2\pi}\right)^{n-1}   I_\eps I\!I_\eps 
\int_{\bbR^{n}}d\beta_{1,n}  \prod_{j=1}^{n-1}(\tilde J\tilde
 g)\left(3\la\eps /4-i\beta_j\right) \prod_{j=1}^{n-1}(\tilde J\tilde
 g)\left(\la\eps/4+i\beta_j\right),
\end{align}
where
\begin{align}
\label{lmI}
I_\eps:=\frac{\ga}{2\pi}
 \int_{\bbR}(\tilde J\tilde
 g)\left(3\la\eps/4-i\beta_0\right) \frac{\tilde g(\la\eps/4+i\beta_0)}{\la\eps/4+i\beta_0-i\om(k)} d\beta_0 
\end{align}
and
\begin{align}
\label{lmII}
&
I\!I_\eps :=\frac{\eps }{2\pi}\int_{\bbT^2}d\ell  d\ell' \int_{\bbR} \tilde g(3\la\eps/4-i\beta_n) \tilde
  g(\la\eps/4+i\beta_n)\notag\\
&
\times\left\{\frac{\bbE\left[\hat\psi(\ell)
  \hat\psi^*(\ell')\right]}{[3\la\eps/4-i\beta_n+i\om(\ell)][\la\eps/4+i\beta_n-i\om(\ell')]}\right.\\
&
\left.+\frac{\bbE\left[\hat\psi^*(\ell) \hat\psi(\ell')\right]}{[3\la\eps/4-i\beta_n-i\om(\ell)][\la\eps/4+i\beta_n+i\om(\ell')]}\right\}d\beta_n. \notag
\end{align}

\subsubsection{The end of the proof of Lemma \ref{lm010312-20}}

\label{sec5.2.5}
Using formula \eqref{090212-20} we conclude, cf \eqref{070212-20n} and
\eqref{Gamma}, that
\begin{equation}
\label{Len}
\lim_{\eps\to0+}{\frak
  L}_\eps^{(n)}=\lim_{\eps\to0+}\bar{\frak
  L}_\eps^{(n)},
\end{equation}
where
\begin{equation}
\label{070212-20nn}
\begin{split}
\bar{\frak
  L}_\eps^{(n)}:=2\int_{\bbR}\int_{\bbT}{\rm Re}\,\tilde D_n^{\eps}\left(\la,k\right)\frac{G^\star(\eta,k)}{\la+i\om'(k)\eta
}d\eta dk.
\end{split}
\end{equation}
Here
$$
\tilde D_n^{\eps}(\la,k):=\frac{\Gamma^{n-1}}{4\mu^n} I_\eps\, I\!I_\eps.
$$
The calculation  of the limit \eqref{Len}   reduces therefore to computing the
limits of $I_\eps$ and $I\!I_\eps$.

\subsubsection*{Computation of $\lim_{\eps\to0+}I_\eps$}

 Since 
$
\tilde g(\la)=1-\ga \tilde J\tilde g(\la)
$
we can write $I_\eps=I_\eps^1+ I_\eps^2$, where
\begin{align*}
&
I_\eps^1:=\frac{\ga}{2\pi}
\int_{\bbR}\frac{(\tilde J\tilde
 g)\left(3\la\eps/4-i\beta_0\right)}{\la\eps/4+i\beta_0-i\om(k)} d\beta_0 \\
&
I_\eps^2:=-\frac{\ga^2}{2\pi}
\int_{\bbR} (\tilde J\tilde
 g)\left(3\la\eps/4-i\beta_0\right) \frac{(\tilde J\tilde
 g) (\la\eps/4+i\beta_0)}{\la\eps/4+i\beta_0-i\om(k)} d\beta_0.
\end{align*}
Using \eqref{CF} we get
\begin{align*}
&
I_\eps^1=\frac{\ga}{2\pi}
\int_{\bbR}\frac{(\tilde J\tilde
 g)\left(3\la\eps/4-i\beta_0\right)}{\la\eps/4+i\beta_0-i\om(k)}d\beta_0 
  =\ga(\tilde J\tilde g) \left(\la\eps-i\om(k)\right).
\end{align*}
Therefore
\begin{equation}
\label{060212-20}
\lim_{\eps\to0+}I_\eps^1= 1- \nu(k).
\end{equation}

On the other hand
$$
\lim_{\eps\to 0} (\tilde J\tilde
 g)\left(3\la\eps/4-i\beta_0\right) (\tilde J\tilde
 g) (\la\eps/4+i\beta_0)=|(\tilde J\tilde
 g)|^2(i\beta_0)
$$
 in any $L^p(\bbR)$, $p\in(1,+\infty)$  and pointwise. Therefore, 
 \begin{align*}
&
\lim_{\eps\to0+} I_\eps^2=-\frac{\ga^2}{2\pi}\lim_{\eps\to0+} \,\left\{
\int_{\bbR}
\frac{|(\tilde J\tilde g)\left(i\beta_0\right)|^2 d\beta_0
  }{\la\eps/4+i\beta_0-i\om(k)}\right\} .
\end{align*}
Since $\textcolor{blue}{j(\beta_0)}:=|(\tilde J\tilde
g)\left(i\beta_0\right)|^2 $ belongs to any
$L^p(\bbR)$ for $p\in[1,+\infty)$, by the multiplier theorem, see e.g. \cite[Corollary of Theorem 3, p. 96]{stein} 
$$
\lim_{\eps\to0+} 
\int_{\bbR}
\frac{j(\beta_0) d\beta_0
  }{\la\eps/4+i\beta_0-i\beta}  =\textcolor{blue}{{\frak j}(\beta)}:=2\pi
\int_{-\infty}^0e^{2\pi i \eta\beta}\hat j(\eta)d\eta,
$$
in the $L^p(\bbR)$ sense, for any $p\in(1,+\infty)$. 
Here
$$
\hat j(\eta):=\int_{\bbR}e^{-2\pi i\eta \beta}j(\beta) d\beta
$$
is the Fourier transform of $j$. 

We have  $\om_+^{-1}(\om_{\rm min})=0$, $\om_+^{-1} (\om_{\rm max})=1/2$.  In the case $\om\in C^{\infty}(\bbT)$:
\begin{equation}\label{opt}
\big(\om_\pm ^{-1}\big) '(w)=\pm (w-\om_{\rm min})^{-1/2}\rho_*(w),~~ w-\om_{\rm min}\ll1,
\end{equation}
and 
\begin{equation}\label{opt-b}
\big(\om_\pm ^{-1}\big) '(w) = \pm (\om_{\rm max}-w)^{-1/2}\rho^*(w),~~\om_{\rm max}-w\ll1,
\end{equation}
with $\rho_*,\rho^*\in C^\infty(\bbT)$ that are strictly positive. 
When $\om$ is not differentiable at $0$ (the acoustic case) 
condition \eqref{opt-b} does not change but  then  
\begin{equation}\label{opt-a}
\big(\om_\pm ^{-1}\big)'(w)=\pm \rho_*(w),~~ w-\om_{\rm min}\ll1.
\end{equation}
In consequence,
\begin{equation}
\label{020312-20}
\lim_{\eps\to0+} I_\eps^2=-\frac{\ga^2}{2\pi}{\frak j}(\om(k))
\end{equation}
in the $L^p(\bbT)$ sense for any $p\in[1,2)$.
We have shown therefore that
\begin{equation}
\label{030312-20}
\lim_{\eps\to0+} I_\eps=I:= 1- \nu(k)-\frac{\ga^2}{2\pi}{\frak j}(\om(k))
\end{equation}
in the $L^p(\bbT)$ sense for any $p\in[1,2)$.
 Since $j$ is real valued we have
\begin{equation}
\label{010312-20}
\textcolor{blue}{\frac{1}{2\pi}{\rm Re}\,{\frak j}(\beta)}=\frac{1}{2}j(\beta)
\end{equation}
and
\textcolor{blue}{
$$
\frac{1}{2\pi}{\rm Re}\,{\frak j}(\om(k))=\frac{1 }{2}|(\tilde J\tilde g)\left(i\om(k)\right)|^2.
$$}
Thus, using the relation 
$$
\ga(\tilde J\tilde g)(\la)=1-\tilde g(\la),
$$
we conclude that
\begin{equation}
\label{030312-20a}
\begin{split}
&{\rm Re}\,I:= 1- {\rm Re}\,\nu(k)-\frac{\ga^2}{2}|(\tilde J\tilde
g)\left(i\om(k)\right)|^2\\
&
=1- {\rm
  Re}\,\nu(k)-\frac{1}{2}|1-\nu(k)|^2=\frac{1}{2}\,\big(1-|\nu(k)|^2\big).
\end{split}
\end{equation}

\subsubsection*{Computation of $\lim_{\eps\to0+}I\!I_\eps $}

We have
$I\!I_\eps= I\!I_\eps^1+ I\!I_\eps^2$, where
\begin{align*}
&
I\!I_\eps^1 :=\frac{\eps }{2\pi}\int_{\bbT^2}d\ell  d\ell'  \int_{\bbR}d\beta_n \frac{\tilde g(3\la\eps/4-i\beta_n) \tilde
  g(\la\eps/4+i\beta_n)\bbE\left[\hat\psi(\ell)
  \hat\psi^*(\ell')\right]}{[3\la\eps/4-i\beta_n+i\om(\ell)][\la\eps/4+i\beta_n-i\om(\ell')]}\\
&
I\!I_\eps^2 := \frac{\eps }{2\pi}\int_{\bbT^2}d\ell  d\ell'  \int_{\bbR}d\beta_n \frac{\tilde g(3\la\eps/4-i\beta_n) \tilde
  g(\la\eps/4+i\beta_n)\bbE\left[\hat\psi^*(\ell) \hat\psi(\ell')\right]}{[3\la\eps/4-i\beta_n-i\om(\ell)][\la\eps/4+i\beta_n+i\om(\ell')]}.
\end{align*}


Changing variables $\eps\beta_n':=\beta_n-\om(\ell')$
we obtain
$$
I\!I_\eps^1 :=\frac{1}{2\pi}\int_{\bbT^2}d\ell  d\ell' \int_{\bbR}d\beta_n \frac{\tilde g(3\la\eps/4-i\eps\beta_n-i\om(\ell)) \tilde
  g(\la\eps/4+i\eps\beta_n+i\om(\ell))\bbE\left[\hat\psi(\ell)
  \hat\psi^*(\ell')\right]}{[3\la/4-i\beta_n+i\eps^{-1}\big(\om(\ell)-\om(\ell')\big)][\la/4+i\beta_n]}.
$$
Therefore
$$
\lim_{\eps\to0+}I\!I_\eps^1 
=\frac{1}{2\pi}\lim_{\eps\to0+}\int_{\bbT^2} \frac{|\nu(\ell)|^2\bbE\left[\hat\psi(\ell)
  \hat\psi^*(\ell')\right]}{\la+i\eps^{-1}\big(\om(\ell)-\om(\ell')\big)}d\ell
d\ell'.
 $$
Changing again variables
$$
\ell=\tilde \ell+\frac{\eps\eta}{2},\quad \ell'=\tilde \ell-\frac{\eps\eta}{2}
$$
we conclude that
\begin{align}
\label{010207-20}
\lim_{\eps\to0+}I\!I_\eps^1=2 \int_{\bbR\times \bbT}
  \frac{|\nu(\ell)|^2\widehat W(0,\eta,\ell)}{\la+i\om'(\ell)\eta}d\eta d\ell.
\end{align}
A similar calculation proves that also
\begin{align}
\label{010207-20a}
\lim_{\eps\to0+}I\!I_\eps^2 =2 \int_{\bbR\times \bbT}
  \frac{|\nu(\ell)|^2\widehat W(0,\eta,\ell)}{\la+i\om'(\ell)\eta}d\eta d\ell.
\end{align}
We conclude therefore 
\begin{align}
\label{010207-20b}
I\!I=\lim_{\eps\to0+}I\!I_\eps=4  \int_{\bbR\times \bbT}
  \frac{|\nu(\ell)|^2\widehat W(0,\eta,\ell)}{\la+i\om'(\ell)\eta}d\eta d\ell.
\end{align}
The right hand side of \eqref{010207-20b} is real valued.
Gathering all the facts proven above we conclude that
\begin{align}
\label{040312-20}
&
\lim_{\eps\to0+}\bar{\frak
  L}_\eps^{(n)}=\frac{\Gamma^{n-1}}{2\mu^n} \int_{\bbR\times \bbT}I\!I{\rm
  Re}\,I\frac{G^\star(\eta,k)}{\la+i\om'(k)\eta}d\eta dk \\
&
=\frac{\Gamma^{n-1}}{\mu^n}\,  \int_{\bbR\times \bbT}\frac{\big(1-|\nu(k)|^2\big)G^\star(\eta,k)}{\la+i\om'(k)\eta}d\eta dk \int_{\bbR\times \bbT} \frac{|\nu(\ell)|^2\widehat
  W(0,\eta',\ell)}{\la+i\om'(\ell)\eta'}d\eta' d\ell.\notag
\end{align}
Combining this with formula \eqref{070212-20nn} we conclude the proof
of  Lemma
\ref{lm010312-20}.\qed

\subsection{Proof of Proposition \ref{prop010412-20}}

\label{sec5.3a}
According to \eqref{exp-wigner-eqt} for any we have
\begin{equation}
\label{exp-wigner-eqt10}
\begin{split}
&
\int_{\bbR\times \bbT}\widehat
   w_\eps(\la,\eta,k)G^*(\eta,k)d\eta dk =\sum_{j=1}^3{\cal W}_j^{(\eps)},\quad\mbox{where}\\
 &
{\cal W}_1^{(\eps)}:=\int_{\bbR\times \bbT}\frac{W_\eps(0,\eta,k)G^*(\eta,k)}{ \la+i\delta_{\eps}\om(k;\eta) }d\eta dk\\
&
{\cal W}_2^{(\eps)}:=\frac{\ga {\frak
    e}_\eps(\la)}{\mu}\int_{\bbR\times \bbT}\frac{G^*(\eta,k)}{ \la+i\delta_{\eps}\om(k;\eta)
   }d\eta dk \\
&
{\cal W}_3^{(\eps)}:=-\frac{\ga}{2}\int_{\bbR\times \bbT} \frac{G^*(\eta,k)}{\la+i\delta_{\eps}\om(k;\eta)}\left[{\frak
    d}_\eps\left(\la,k-\frac{\eps \eta}{2}\right)+{\frak
    d}_\eps^\star\left(\la,k+\frac{\eps \eta}{2}\right)\right]d\eta dk.
\end{split}
\end{equation}
It is easy to see that the limit of ${\cal W}_1^{(\eps)}$, as
$\eps\to0+$, corresponds to the first term in the right hand side of
\eqref{010412-20}. Using Proposition \ref{prop010212-20} we conclude
that the limit of ${\cal W}_2^{(\eps)}$ matches  the second term
there. Finally
$
{\cal W}_3^{(\eps)}=-\frac{\ga}{2}\sum_{n=0}^{+\infty}{\frak L}_{\eps}^{(n)}
$
and the respective limit is a consequence of Lemmas \ref{lem-feb1504},
\ref{lem-feb1502} and \ref{lm010312-20}. This ends the proof of the proposition.\qed

\subsection{The end of the proof of Theorem \ref{main:thm}}

\label{sec:nul-init-cond}

Using the equality \eqref{hwigner-12} and the results of Proposition
\ref{prop010112-20} (for $\mu>1/2$), Lemma \ref{lm010112-20} (for
$\mu=1/2$) and Proposition \ref{prop010112-20}, together with formula
\eqref{010412-20} we conclude that for any $\la\in\mathbb C_+$ the Laplace-Fourier-Wigner functions
$\widehat{ w}_\eps(\la,\eta,k)$ converge, as $\eps\to0+$, in ${\cal A}'$, in the
$\star$-weak topology to 
\begin{align}
\label{010407-20}
&
\widehat{ w}(\la,\eta,k)=\frac{\widehat
  W(0,\eta,k)}{\la+i\om'(k)\eta}+
\frac{  \ga T|\nu(k)|^2}{(1-\Gamma/\mu)\la(\la+i\om'(k)\eta)}\left(1-\frac{1}{2\mu}\right)
 \\
&
+ 
\frac{\ga |\nu(k)|^2}{2\mu[\la+i\om'(k)\eta](1-\Gamma/\mu)} 
  \int_{\bbR\times\bbT} \frac{|\nu(\ell)|^2\widehat
  W(0,\eta',\ell)}{\la+i\om'(\ell)\eta'}d\eta' d\ell
\notag 
-\frac{\ga {\rm Re} [\nu(k)]}{\la+i\om'(k)\eta} \int_{\bbR}
 \frac{\widehat W(0,\eta',k)}{\la+i\om'(k)\eta'} 
 d\eta'
\\
&
+\frac{\gamma \fgeeszett(k)}{4(\la+i\om'(k)\eta)}\int_{\bbR }\frac{\widehat
  W(0,\eta',k)d\eta' }{\la+i\om'(k)\eta'} 
+\frac{\gamma \fgeeszett(k)}{4(\la+i\om'(k)\eta)}\int_{\bbR}\frac{\widehat
  W(0,\eta',-k)d\eta' }{\la-i\om'(k)\eta'} 
. \notag
\end{align}
Inverting both the Laplace transform in $t$ and Fourier transform in
$x$ we obtain \eqref{010304}, which ends the proof of the theorem.\qed


\section{Proofs of Lemmas \ref{lm011811-20} and \ref{lm013011-20}}

\label{sec10}

\subsection{Proof of Lemma \ref{lm011811-20}}

\label{sec10.1}

We have
\begin{equation}
\label{020512-20}
 \tilde J(\la)
=G\big(\la\big)+H\big(\la\big) ,
\end{equation}
where
\begin{equation}
\label{G-H}
G(\la):=\frac12\int_{\bbT_+}\frac{ d\ell}{\la+i\om(\ell)}
\qquad
H(\la):= \frac 12\int_{\bbT_+}\frac{
  d\ell}{\la-i\om(\ell)}.
\end{equation}
Thanks to \eqref{020512-20} and \eqref{012410} we conclude that
\begin{equation}
\label{010512-20}
|(\tilde g\tilde J)(\la)|\le \frac{1}{|\la|-\om_{\rm max}},\quad |\la|>\om_{\rm
   max},\,{\rm Re}\,\la>0.
\end{equation}
On the other hand, thanks to \eqref{012410} and \eqref{tgJ}, we have also
\begin{equation}
\label{010512-20a}
 |(\tilde g\tilde J)(\la)|\le \frac{2}{\ga},\quad\,{\rm Re}\,\la>0.
\end{equation}
As a result $\tilde g\tilde J\in H^p(\mathbb C_+)$ for any
$p\in(1,+\infty)$. The limits in \eqref{030512-20} and \eqref{nuk} can
be substantiated by the results of Sections A and  B of Chapter 6 of  \cite{koosis}.

{Recall that $\omega_+^{-1}(\cdot)$ is the inverse of the restriction $\om_{[0,1/2]} $.}   From \eqref{G-H} we get
\begin{align*}
G(\eps+i\om(k))=\frac12\int_{\omega_{\min}}^{\omega_{\max}}\frac{dv}{\omega'(\omega_+^{-1}(v))[\eps+i(v+\om(k))]}
\end{align*}
To simplify assume that $k\in[0,1/2]$. It is clear that
\begin{align*}
\lim_{\eps\to0+}G(\eps+i\om(k))=G(i\om(k))=-\frac i2\int_{\omega_{\min}}^{\omega_{\max}}\frac{dv}{\omega'(\omega_+^{-1}(v))(v+\om(k))}
\end{align*}
and there exists $C>0$ such that 
\begin{equation}
\label{040512-20}
\Big|G(\eps+i\om(k))-G(i\om(k))\Big|\le C\eps,\quad k\in \Om_*^{(\delta)},
\end{equation}
{where $\Om_*^{(\delta)}:=[k\in\bbT:\,{\rm dist}\big(k, \Om_* \big)\ge\delta]$.}
{Concerning $H(\cdot)$ we have}
\begin{align*}
H(\eps+i\om(k))=\frac12\int_{\omega_{\min}}^{\omega_{\max}}\frac{dv}{\omega'(\omega_+^{-1}(v))[\eps+i(\om(k)-v)]}
\end{align*}
{A simple calculation leads to}
\begin{align*}
&H(i\om(k)):=\lim_{\eps\to0+}H (\eps+i\om(k))=\frac1{2
 \omega'(k)}\left[\pi+i\log\left(\frac{\om_{\rm
        max}-\om(k)}{\om(k)-\om_{\rm min}}\right)\right]\\
&
+
\frac i2\int_{\omega_{\min}}^{\omega_{\max}}\frac{[\om'(k) -\omega'(\omega_+^{-1}(v))] dv}{\omega'(\omega_+^{-1}(v))\om'(k)(\om(k)-v)}.
\end{align*}
{Since $\om'(\cdot)$ is Lipschitz the integral in the right hand side
makes sense. A straightforward calculation implies the existence of  $C>0$ such that }
\begin{equation}
\label{050512-20}
\Big|H(\eps+i\om(k))-H(i\om(k))\Big|\le C\eps,\quad k\in \Om_*^{(\delta)}.
\end{equation}
From \eqref{040512-20} and \eqref{050512-20} we conclude
\eqref{060512-20}. In addition we infer also the continuity of $\nu$
on $\bbT\setminus \Om_*$.
\qed

\subsection{Proof of Lemma \ref{lm013011-20}}

\label{sec10.2}


For a given $f\in L^1(\bbR)$ such that $f\ge0$ a.e. we let
\begin{equation}
\label{Mz}
M(z):=\int_{\bbR}\frac{f(\al)d\al}{z+i\al},\quad z\in\mathbb C_+.
\end{equation}
The function is  holomorphic and ${\rm
  Re}\,M(z)>0$ for  $z\in \mathbb C_+$. In addition, for any $\rho>0$ we have
\begin{equation}
\label{Mza}
M(\rho+i\beta):=\int_{\bbR}\frac{\rho f(\al)d\al}{\rho^2+(\beta+\al)^2}-i\int_{\bbR}\frac{(\beta+\al) f(\al)d\al}{\rho^2+(\beta+\al)^2},\quad \beta\in\bbR.
\end{equation}
Suppose also that $f\in L^p(\bbR)$ for some $p>1$. By \cite[Corollary of
Theorem 3, p. 96]{stein} we conclude that
\begin{equation}
\label{Mzb}
M_+(\beta):=\lim_{\rho\to0+}M(\rho+i\beta)=\pi f(-\beta)-i{\cal H}[f](\beta),\quad\beta\in\bbR,
\end{equation}
where 
\begin{equation}
\label{Mzc}
{\cal H}[f](\beta):=\lim_{\rho\to0+}\int_{\bbR}\frac{(\beta+\al) f(\al)d\al}{\rho^2+(\beta+\al)^2},\quad\beta\in\bbR,
\end{equation}
and the limits in  \eqref{Mzb} and \eqref{Mzc} are understood in the
$L^p$ sense.

We shall prove the following result.
\begin{prop}
\label{prop012601-21}
Suppose that $f\in L^1(\bbR)\cap  L^p(\bbR)$ for some $p>1$  and $f\ge0$ a.e. Then, for
any $\ga>0$ the
following identity holds
\begin{equation}
\label{basid}
\frac{\ga}{2\pi}\int_{\bbR}\frac{|M_+(\beta)|^2d\beta}{|1+\ga M_+(\beta)|^2}+\frac{1}{2}\int_{\bbR}\frac{f(-\beta)d\beta}{|1+\ga M_+(\beta)|^2}=\frac{1}{2}\int_{\bbR}f(\beta)d\beta.
\end{equation}
\end{prop}
Before proving the proposition, which we are going to do momentarily, let
us first apply it to show how, with its help, to finish the proof of
Lemma \ref{lm013011-20}.

\subsubsection{Proof of Lemma  \ref{lm013011-20}}
From \eqref{eq:2} we get
\begin{equation}
\label{012501-21}
\tilde
J(\la)=\int_0^{1/2}\frac{dk}{\la+i\om(k)}+\int_0^{1/2}\frac{dk}{\la-i\om(k)}=\int_{\bbR}\frac{f_*(v)dv}{\la+iv},
\end{equation}
where
\begin{equation}
\label{012501-21a}
f_*(v):=\frac{1_{[\om_{\rm min},\om_{\rm
      max}]}(|v|)}{\om'\big(\om_+^{-1}(|v|)\big)},\quad v\in\bbR.
\end{equation}
Recalling that 
$$\om'\big(\om_+^{-1}(v)\big)\sim (\om_{\rm
  max}-v)^{1/2},\quad  \om_{\rm
  max}-v \ll 1,
$$
see   \eqref{opt}, and 
$$
\om'\big(\om_+^{-1}(v)\big)\sim (v-\om_{\rm
  min})^{1/2},\quad v-\om_{\rm
  min}\ll 1
$$ in the optical case  (see \eqref{opt-b}), and
$|\om'\big(\om_+^{-1}(v)\big)|\sim 1$, $v\ll1$ in the acoustic one 
we conclude that $f_*\in L^p(\bbR)$ for any $p\in[1,2)$ and
$\int_{\bbR}f_*(v)dv=1$. It is easy to see from \eqref{012501-21} and
\eqref{012501-21a} that
\begin{equation}
\label{022501-21}
J^\star(\la)=J(\la^\star),\quad \la\in\mathbb C_+.
\end{equation}
Recall that   $\tilde J(i\om(k))=\lim_{\eps\to0+}\tilde J(\eps+i\om(k))$, cf
\eqref{030512-20},  
therefore
\begin{align*}
&\int_{\bbT}|\nu(\ell)|^2 d\ell=\int_{\bbT}\frac{d\ell}{|1+\ga\tilde
  J(i\om(\ell))|^2}\\
&
=\int_{0}^{1/2}\frac{d\ell}{|1+\ga\tilde
  J(i\om(\ell))|^2}+\int_{0}^{1/2}\frac{d\ell}{|1+\ga\tilde
  J^\star(i\om(\ell))|^2}\\
&
=\int_{0}^{1/2}\frac{d\ell}{|1+\ga\tilde
  J(i\om(\ell))|^2}+\int_{0}^{1/2}\frac{d\ell}{|1+\ga\tilde
  J(-i\om(\ell))|^2}=
\int_{\bbR}\frac{f_*(v)dv}{|1+\ga\tilde J(iv)|^2}.
\end{align*}
Formula \eqref{022712-20} is then a direct consequence of \eqref{basid}.
Equality  \eqref{022712-20} is in fact equivalent with 
\begin{align}
\label{012712-20}
1=\frac{
  1}{2(1-\Gamma)}
+\frac{1}{2(1-\Gamma)}
\int_{\bbT}|\nu(\ell)|^2 d\ell,
\end{align}
which in turn yields
\eqref{fund-ident2}.\qed

\subsubsection{Proof of Proposition \ref{prop012601-21}}

\bigskip

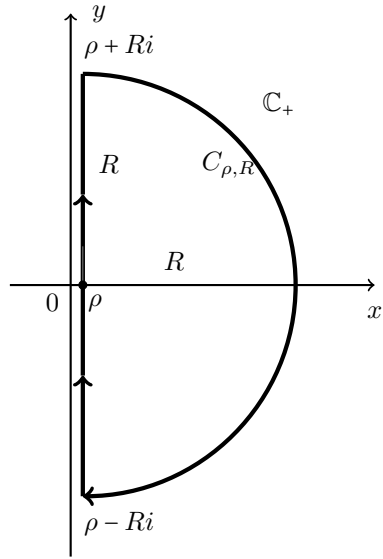
\begin{figure}[ht]

\begin{center}
\begin{tikzpicture}[scale=0.8]

\draw [thick,->]  (-1,0) -- (5,0);
\node  [below] at (-0.3,0) {$0$};
\node  [below] at (0.4,0) {$\rho$};

\node  [below] at (5,-0.2) {$x$};

\node  [right] at (0.2,4.5) {$y$};

\draw [thick,->]  (0,-4.5) -- (0,4.5);

\node  [below] at (0.8,-3.6) {$\rho-Ri$};
\node  [above] at (0.8,3.6) {$\rho+Ri$};

\node  [above] at (1.7,0.1) {$R$};

\draw [ultra thick,->]  (0.2,-3.5) -- (0.2,-1.5);
\draw [ultra thick,->]  (0.2,-1.5) -- (0.2,1.5);
\draw [ultra thick,-] (0.2,1.5) -- (0.2,3.5);

\node [fill, draw, circle, minimum width=3pt, inner sep=0pt,
pin={[outer sep=2pt]}] at (0.2,0) {};

 \node  [right] at (0.3,2) {$R$};

\draw[ultra thick, ->] (0.2,3.5) arc (90:-90:3.5);

\node  [right] at (2,2) {$C_{\rho,R}$};
\node  [right] at (3,3) {$\mathbb C_+$};

 \end{tikzpicture}
\caption{Contour of integration} 
\label{fig1}
\end{center}
\end{figure}

Suppose that $\rho,R>0$. Consider the contour $C_{\rho,R}$, cf Figure \ref{fig1}, made of the
line segment from $\rho-Ri$ to $\rho+Ri$ and the semicircle centered
at $\rho$ of radius $R$, oriented clockwise.
Since $M(z)$ is analytic in $\mathbb C_+$ we have
\begin{equation}
\label{012601-21}
\int_{C_{\rho,R}}\frac{M(z)dz}{1+\ga M(z)}=0.
\end{equation}
The above equality yields
\begin{equation}
\label{012601-21aa}
\int_{-R}^R\frac{M(\rho+i\beta)d\beta}{1+\ga M(\rho+i\beta)}=\int_{-\pi/2}^{\pi/2}\frac{M(\rho+Re^{i\theta})Re^{i\theta}d\theta}{1+\ga M(\rho+Re^{i\theta})}.
\end{equation}
Letting first $\rho\to0+$ and then $R\to+\infty$, in this order, we
conclude, thanks to the definition of $M(z)$ and the fact that the
expression under the integral is bounded, that 
\begin{equation}
\label{012601-21a}
\lim_{R\to+\infty}\int_{-R}^R\frac{M_+(\beta)d\beta}{1+\ga M_+(\beta)}=\pi\int_{\bbR}f(\beta)d\beta.
\end{equation}
Taking complex conjugation on both sides
\begin{equation} 
\label{012601-21b}
\lim_{R\to+\infty}\int_{-R}^R\frac{M_+^\star(\beta)d\beta}{1+\ga M_+^\star(\beta)}=\pi\int_{\bbR}f(\beta)d\beta.
\end{equation}
Adding \eqref{012601-21a} and \eqref{012601-21b}
sideways, and using \eqref{Mzb} we get
\begin{equation}
\label{012601-21c}
\begin{split}
&2\ga\lim_{R\to+\infty}\int_{-R}^R\frac{|M_+(\beta)|^2d\beta}{|1+\ga
  M_+(\beta)|^2}+2\pi\lim_{R\to+\infty}\int_{-R}^R\frac{f(-\beta)d\beta}{|1+\ga
  M_+(\beta)|^2}\\
&
=2\pi\int_{\bbR}f(\beta)d\beta.
\end{split}
\end{equation}
This ends the proof of the proposition. \qed

{\small

\end{document}